\documentclass[prl,aps,twocolumn,superscriptaddress,amsmath,longbibliography]{revtex4-2}
\usepackage{graphicx,color,hyperref,xcolor}
\usepackage{soul}
\usepackage{amsfonts}
\usepackage{amssymb}
\usepackage{amsmath}
\usepackage{amsthm}
\usepackage{bm}
\usepackage{braket}
\usepackage{enumerate}

\usepackage{CJKutf8}

\begin{document}
\begin{CJK}{UTF8}{gbsn}

\title{Optical engineering and detection of magnetism in moir\'e semiconductors}

\author{Tsung-Sheng Huang}
\affiliation{Joint Quantum Institute, University of Maryland, College Park, MD 20742, USA}

\author{Andrey Grankin}
\affiliation{Joint Quantum Institute, University of Maryland, College Park, MD 20742, USA}

\author{Yu-Xin Wang (王语馨)}
\affiliation{Joint Center for Quantum Information and Computer Science, University of Maryland, College Park, MD 20742, USA}

\author{Mohammad Hafezi}
\affiliation{Joint Quantum Institute, University of Maryland, College Park, MD 20742, USA}
\affiliation{Joint Center for Quantum Information and Computer Science, University of Maryland, College Park, MD 20742, USA}

\date{\today}

\begin{abstract}

We present a general framework for optically inducing, controlling, and probing spin states in moiré systems. In particular, we demonstrate that applying Raman optical drives to moiré transition metal dichalcogenide bilayers can realize a class of spin models, with magnetic interactions tunable via the optical parameters. The resulting interaction anisotropy, controlled by the polarizations of the drives, enables access to magnetic states that are inaccessible in undriven moiré bilayers. Furthermore, we establish direct connections between the resulting spin correlations and experimentally observable optical signals. Our work paves the way for future studies on the optical control and detection on strongly correlated quantum systems.

\end{abstract}

\maketitle
\end{CJK}

\textit{Introduction.} --- Ultracold atoms in optical lattices have proven to be an exceptionally powerful platform for exploring a wide range of models, from generalized Heisenberg spin and Fermi-Hubbard models~\cite{bloch2008many} to various topological systems~\cite{cooper2019topological}. These advances have been made possible through the optical engineering of diverse Hamiltonians, including tunneling and interaction terms~\cite{feshbach1962unified,duan2003controlling}.

Motivated by these developments, an intriguing question arises: Can similar optical control be applied to a many-electron system and induce, modify and probe correlated phenomena~\cite{bloch2022strongly}? 
Specifically, is it possible to optically induce and probe magnetism, by building upon a microscopic model of light-matter interaction?  


Moiré transition metal dichalcogenide (TMD) bilayers~\cite{mak2022semiconductor,du2023moire,kennes2021moire}, which offers a rich landscape of strong correlations~\cite{kennes2021moire}, provide a promising platform for this investigation. 
The nonlinearities of electrons~\cite{wu2018hubbard,pan2020quantum} and excitons~\cite{gotting2022moire,huang2024nonbosonic,song2024electrically,song2024microscopic} in these systems enable the emergence of generalized Fermi-Hubbard~\cite{wu2018hubbard,tang2020simulation,wang2020correlated} and generalized Bose-Hubbard~\cite{park2023dipole,xiong2023correlated,gao2024excitonic} physics, respectively, and their coexistence allows for the realization of Bose-Fermi mixtures~\cite{miao2021strong,mhenni2024gate,upadhyay2024giant}.
Importantly, spin-resolved observables from these correlated phenomena, where spins are locked to valley pseudospins via spin-orbit coupling~\cite{xiao2012coupled}, may be optically probed~\cite{salvador2022optical} or controlled by exploiting the selectivity in TMD heterostructures: 
direct transitions at valley pseudospin $\tau = +$ ($\tau = -$) primarily couple to photons with circular polarization $\sigma_+$ ($\sigma_-$)~\cite{yu2017moire}.

\begin{figure}[t]
\centering
\includegraphics[width=\columnwidth]{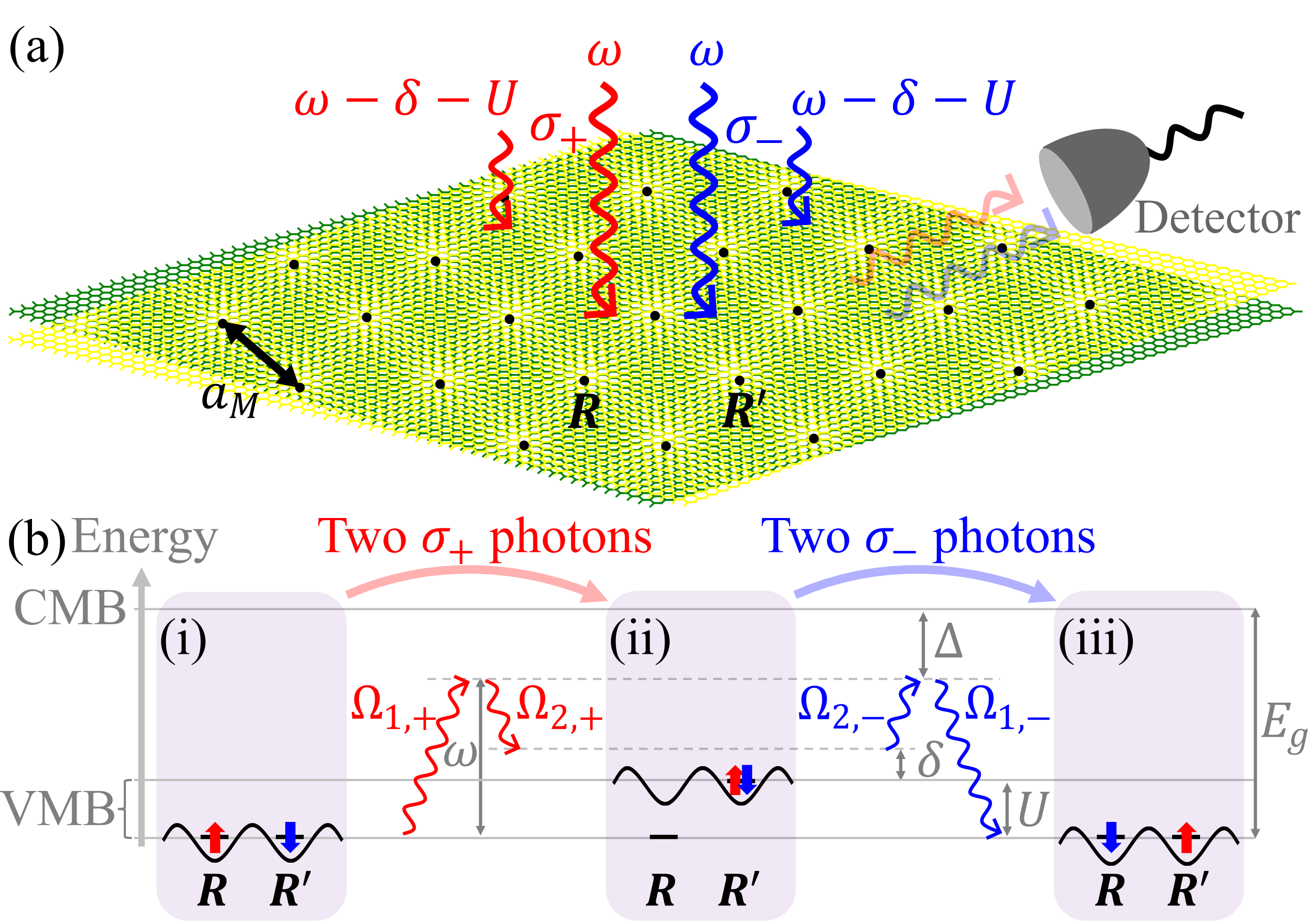}
\caption{
Illustration of the system.
(a) A moir\'e heterobilayer with superlattice spacing $a_M$ driven by two external light sources at distinct frequencies $\omega$ and $\omega-\delta-U$.
Emitted photons are collected by the detector.
$\bm{R}$ and $\bm{R}'$ denote moir\'e sites.
(b) The two-photon Raman processes inducing spin exchanges in the valence moir\'e band (VMB).
In the presence of a charge gap $U$, VMB splits into upper and lower Hubbard bands.
Up- and down-arrows located at the moir\'e sites denote VMB electrons at $\tau=+$ and $-$ valleys, respectively. 
A photon at frequency $\omega$ enables transition from the lower Hubbard band to a virtual state detuned from the conduction moir\'e band (CMB) by $\Delta = E_g - \omega$, and a photon at $\omega-\delta-U$ couples such virtual state to another that is detuned from the upper Hubbard band by $\delta$.
$\Omega_{\lambda,\tau}$ with $\lambda\in\{1,2\}$ and $\tau\in\{\pm\}$ labels the components of the drives.
Black curves indicate the moir\'e landscape for VMB.
}
\label{Fig_1}
\end{figure}

In this work, we propose how various spin models can be optically engineered in moiré TMDs, as sketched in Fig.\ref{Fig_1}(a). 
Specifically, we focus on hole-doped heterobilayers at half-filling, where the Fermi-Hubbard gap $U$ emerges within the first valence moiré band (VMB)\cite{wu2018hubbard}.
Two light fields with frequencies $\omega$ and $\omega-\delta-U$ induce Raman transitions between the lower and upper Hubbard bands within VMB and the first conduction moiré band (CMB). 
For instance, as shown in Fig.~\ref{Fig_1}(b), a $\sigma_+$-polarized photon with frequency $\omega$ can virtually excite a $\tau = +$ electron from VMB to CMB. 
The charge can then return via the same photon, causing an AC Stark shift, or create a VMB double occupancy at another site with a $\omega - \delta - U$ drive at the same polarization.
Notably, this optically induced tunneling occurs only if the initial and final site spins do not form a triplet state. 
The double occupancy can be eliminated via the reverse process, driven by either the same-($\sigma_+$) or opposite ($\sigma_-$)-polarized photons. 
While the former induces $\hat{S}_{\bm{R}}^z$ and $\hat{S}_{\bm{R}}^z\hat{S}_{\bm{R}'}^z$ contributions, the latter yields $\hat{S}_{\bm{R}}^x\hat{S}_{\bm{R}'}^x+\hat{S}_{\bm{R}}^y\hat{S}_{\bm{R}'}^y$, where $\hat{S}_{\bm{R}}^j$ represents the VMB spins at supersite $\bm{R}$ and orientation $j\in\{x,y,z\}$.

We show that this scenario realizes a triangular lattice $XXZ$ model~\cite{yamamoto2014quantum,sellmann2015phase}, with spin-spin interactions tunable via external drives. 
Their strength can be enhanced by adjusting drive intensities and frequencies, making magnetism more robust and experimentally accessible.
Moreover, the interaction anisotropy is tunable via the drive polarizations, enabling previously unexplored magnetic phases in moiré TMDs. 
This highlights how optical control can further advance the prospect of quantum simulation in moiré TMDs.
Finally, we show that these driven magnetic correlations can be probed through optical observables, specifically the intensity of Stokes sideband emission~\cite{gangopadhyay1992spectral}.
This method provides a more direct signature of magnetic phases compared to the conventional reflective magnetic circular dichroism measurement and Curie-Weiss analysis~\cite{tang2020simulation,anderson2023programming,ciorciaro2023kinetic}, which does not reveal interaction anisotropy.


\textit{Two-band model.} --- 
We investigate a TMD heterobilayer where electronic interaction in the highest valence moiré band (VMB) is described by an on-site Hubbard repulsion~\cite{wu2018hubbard}. 
For simplicity, we focus on optical transitions between the VMB and the lowest conduction moiré band (CMB), assuming others are detuned and hence negligible. 
To exploit valley-polarization selectivity~\cite{yu2017moire}, we further consider \textit{intralayer} VMB and CMB. 
Such a configuration can be realized in heterobilayers with type-I band alignment, such as MoSe\textsubscript{2}/WS\textsubscript{2}~\cite{kang2013band}.

With these considerations, we construct the two-band model $\hat{H}$:
\begin{equation}
\label{eq:starting_model}
\hat{H} 
= 
\hat{U}
+
E_g
\sum_{\bm{L},\tau}
\hat{c}_{\bm{L},\tau}^\dagger 
\hat{c}_{\bm{L},\tau}
+ 
\hat{V}_{\mathrm{LM}} 
+ 
\hat{H}_{\mathrm{ph}}
,
\end{equation}
where $\hat{U} = U\sum_{\bm{R}}\hat{n}_{\bm{R},+}\hat{n}_{\bm{R},-}$ with $\hat{n}_{\bm{R},\tau} = \hat{v}_{\bm{R},\tau}^\dagger \hat{v}_{\bm{R},\tau}$ denoting the number operators of electrons in VMB and $U$ denoting their Hubbard interaction.
$\tau\in\{\pm\}$ labels the valley pseudospin.
$\{\bm{L}\}$ and $\{\bm{R}\}$ are the set of triangular superlattice sites of CMB and VMB electrons, respectively, which may either coincide or be shifted, depending on the material and stacking configuration~\cite{guo2020shedding,naik2022intralayer}.
$E_g$ represents the bandgap.
The Hamiltonian for free electromagnetic field $\hat{H}_{\mathrm{ph}}$ is later eliminated by going to the interaction picture, and accordingly, the electric field operator becomes time dependent, which we express as $\hat{\bm{E}}(t)=\frac{1}{d}\sum_\tau \Omega_\tau(t)\bm{e}_\tau+\delta \hat{\bm{E}}$.
Here $d$ is the transition dipole associated to the atomic orbitals of CMB and VMB, $\Omega_\tau(t)$ represents the component of (spatially homogeneous) classical drive at polarization $\bm{e}_\tau = \frac{\bm{e}_x+i\tau\bm{e}_y}{\sqrt{2}}$ (with $\bm{e}_x$ and $\bm{e}_y$ being Cartesian unit vectors in the plane), and $\delta \hat{\bm{E}}$ captures the quantum fluctuations. 
Note that we will suppress $\delta \hat{\bm{E}}$ when deriving effective Hamilonians but reintroduce it when analyzing optical detection~\footnote{Dropping the contributions of the fluctuations in the effective Hamiltonian is valid if the drives are much stronger than the spontaneous decay rate of optical excitations. In contrast, the fluctuations would yield the main contribution of the scattered photons, and need to be included for describing optical detection.}.
The drives contain two frequencies $\omega$ and $\omega - \delta - U$:
\begin{equation}
\Omega_\tau(t)
=
e^{-i\omega t}
\left[
\Omega_{1,\tau}
+
e^{i(\delta+U)t}
\Omega_{2,\tau}
\right]
,
\end{equation}
with $\Omega_{\lambda,\tau}$ labeling the drive component at mode $\lambda\in\{1,2\}$.
The classical drives couple to the electrons through the following light-matter interactions (in the interaction picture) with valley-polarization selectivity~\cite{yu2017moire}:
\begin{equation}
\label{eq:V_LM}
\hat{V}_{\mathrm{LM}}(t)
\simeq
-
\sum_{\tau}
\sum_{\bm{L},\bm{R}}
\Omega_\tau(t)
\eta_{\bm{L},\bm{R}}
\hat{c}_{\bm{L},\tau}^\dagger
\hat{v}_{\bm{R},\tau}
+
\mathrm{H.c.}
,
\end{equation}
where $\eta_{\bm{L},\bm{R}}$ represents the overlap between the moir\'e-Wannier wavefunctions of CMB and VMB electrons centered at $\bm{L}$ and $\bm{R}$, respectively~\cite{Supplement}.
Notably, the (intrinsic) tunneling of CMB and VMB electrons is ignored in Eq.~\eqref{eq:starting_model}. 
The former is justified, as we later focus on a detuned regime where the CMB is only virtually accessed, while the latter is weaker than a similar term induced by $\hat{V}_{\mathrm{LM}}(t)$, as we demonstrate below.

\begin{figure}[t]
\centering
\includegraphics[width=\columnwidth]{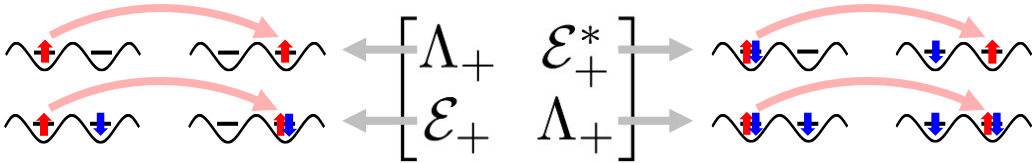}
\caption{
Illustration of optically driven density-dependent tunneling Eq.~\eqref{eq:t_operator} with $\tau = +$.
Notation follows Fig.~\ref{Fig_1}.
}
\label{Fig_2}
\end{figure}

\textit{Optically-induced tunneling.} ---
We proceed to derive an effective model for VMB electrons from Eq.~\eqref{eq:starting_model} in a regime where the optical drives are far off-resonant from $E_g$.
Specifically, we consider the situation where the Raman detuning $\Delta = E_g - \omega$ and the two-photon detuning $\delta$ satisfy the following energy hierarchy:
\begin{equation}
\label{eq:energy_hierarchy}
|\Delta|
\gg
U
\gg
|\delta|
\gg
\left|
\frac{\Omega_{\lambda,\tau}\Omega_{\lambda',\tau}}{\Delta}
\right|
.
\end{equation}
Within these conditions, applying standard perturbation theory to Eq.~\eqref{eq:starting_model} leads to the following effective model:~\cite{auerbach2012interacting}:
\begin{equation}
\label{eq:H_v}
\begin{aligned}
\hat{H}_{v}
&=
\hat{U}
-
\xi_0
\sum_{\bm{R},\tau}
\Lambda_\tau
\hat{n}_{\bm{R},\tau}
-
\xi_1
\sum_{\tau}
\sum_{\langle\bm{R},\bm{R}'\rangle}
\hat{t}_{\bm{R},\bm{R}'}^\tau
\hat{v}_{\bm{R},\tau}^\dagger
\hat{v}_{\bm{R}',\tau}
, 
\end{aligned}
\end{equation}
where $\xi_0 = \sum_{\bm{L}}|\eta_{\bm{L},\bm{R}}|^2$, and
$\Lambda_\tau = \frac{1}{\Delta} \sum_{\lambda = 1}^2 |\Omega_{\lambda,\tau}|^2$ is the AC stark shift of the process where a VMB electron is virtually excited and then returns to its original site.
$\xi_1 = \sum_{\bm{L}} \eta_{\bm{L},\bm{R}}^\ast \eta_{\bm{L},\bm{R}+\bm{a}_M}$ (with $\bm{a}_M$ denoting the moir\'e lattice vector) represents the nearest-neighbor moiré-Wannier function overlap.
Similar long-range terms are omitted in Eq.~\eqref{eq:H_v} as they decay exponentially with separation and are expected to be negligible relative to $\xi_1$.
$\hat{t}_{\bm{R},\bm{R}'}^\tau$ is the optically induced nearest-neighbor density-dependent tunneling~\cite{dutta2015non} with the following expression:
\begin{equation}
\label{eq:t_operator}
\hat{t}_{\bm{R},\bm{R}'}^\tau
=
\begin{bmatrix}
1 - \hat{n}_{\bm{R},-\tau}
\\
\hat{n}_{\bm{R},-\tau}
\end{bmatrix}^\dagger
\begin{bmatrix}
\Lambda_\tau
&
\mathcal{E}_\tau^\ast
\\
\mathcal{E}_\tau
&
\Lambda_\tau
\end{bmatrix}
\begin{bmatrix}
1 - \hat{n}_{\bm{R}',-\tau}
\\
\hat{n}_{\bm{R}',-\tau}
\end{bmatrix}
,
\end{equation}
where $\mathcal{E}_\tau \equiv \frac{\Omega_{2,\tau}^\ast\Omega_{1,\tau}}{\Delta}$ is the effective drive between the lower and upper Hubbard bands from VMB.
The matrix structure in Eq.~\eqref{eq:t_operator} captures the tunneling rates for various processes, see Fig.~\ref{Fig_2}. 
Specifically, the diagonal terms correspond to hoppings that preserve the number of double occupancies, while the off-diagonal sectors describe tunnelings that alter this number.
Finally, these hopping integrals are $\tau$-dependent and their valley contrast can be tuned by the polarization of external drives.
This contrast gives rise to anisotropy in spin-spin interactions, as discussed below.

\textit{Half-filling.} --- When the doping concentration, controlled by the gate voltage, is tuned to one VMB electron per moiré unit cell, we can apply an additional perturbation step based on the energy hierarchy in Eq.~\eqref{eq:energy_hierarchy} to project Eq.~\eqref{eq:H_v} onto the subspace where there is exactly one VMB electron per moiré unit cell~\cite{auerbach2012interacting}.
This procedure yields the following effective spin model: $\hat{H}_{S} = - B_L \sum_{\bm{R}} \hat{S}_{\bm{R}}^z + \hat{H}_{J}$, where $B_L \simeq \xi_0\sum_{\tau} \tau \Lambda_\tau$ is the optically induced Zeeman splitting arising from the AC Stark shift valley contrast, and $\hat{H}_{J}$ captures the following spin-spin interaction of $XXZ$ form (keeping only nearest-neighbor terms):
\begin{equation}
\label{eq:XXZ}
\hat{H}_{J}
=
J
\sum_{\langle\bm{R},\bm{R}'\rangle}
\left[
\hat{S}_{\bm{R}}^z
\hat{S}_{\bm{R}'}^z
+
\alpha
(
\hat{S}_{\bm{R}}^x
\hat{S}_{\bm{R}'}^x
+
\hat{S}_{\bm{R}}^y
\hat{S}_{\bm{R}'}^y
)
\right]
,
\end{equation}
where $\hat{S}_{\bm{R}}^j
=
\frac{1}{2}
\sum_{\tau,\tau'}
\hat{v}_{\bm{R},\tau}^\dagger
\sigma_{\tau,\tau'}^j
\hat{v}_{\bm{R},\tau'}
,\;
\forall j\in\{x,y,z\}$ are the spin operators of VMB electrons with  $\sigma_{\tau,\tau'}^j$ being Pauli matrix elements.
The Ising coupling $J$ and the anisotropy in the spin-spin interaction $\alpha$ have the following expressions, respectively:
\begin{equation}
\label{eq:Jz_and_alpha}
J 
\simeq 
- \frac{2|\xi_1|^2}{\delta} 
\sum_\tau 
|\mathcal{E}_\tau|^2
,\quad
\alpha
\simeq
\frac{2|\mathcal{E}_+\mathcal{E}_-| }{|\mathcal{E}_+|^2 + |\mathcal{E}_-|^2}
\cos\phi_r
.
\end{equation}
The sign of $J$ depends on $\delta$, and its expression suggests that $XXZ$ magnetism becomes more robust with stronger drives or smaller detunings.
In contrast, $\alpha$ is controlled by the polarization parameters $|\mathcal{E}_+/\mathcal{E}_-|$ and $\phi_r = \arg(\Omega_{1,-}\Omega_{2,+}\Omega_{2,-}^\ast\Omega_{1,+}^\ast)$. 
Note that $|\alpha|\leq1$, with $\alpha=1$ (Heisenberg model) and $\alpha=-1$ realized when the two drives are linearly polarized along parallel and orthogonal directions, respectively.
This restriction implies that while the Ising and Heisenberg models are included in Eq.~\eqref{eq:XXZ}, the $XY$ model is not.

The form of Eq.~\eqref{eq:XXZ} can be justified using the symmetries of the light-matter coupling in Eq.~\eqref{eq:V_LM}. 
Specifically, $\hat{V}_{\mathrm{LM}}$ preserves the total out-of-plane magnetization, given by $\sum_{\bm{L},\tau} \tau\hat{c}_{\bm{L},\tau}^\dagger\hat{c}_{\bm{L},\tau} +\sum_{\bm{R},\tau}\tau \hat{n}_{\bm{R},\tau}$ but does not generally conserve the total in-plane magnetization.  
After projecting out CMB states and double occupancies in VMB, the effective Hamiltonian preserves $\sum_{\bm{R}}\hat{S}_{\bm{R}}^z$ but not necessarily $\sum_{\bm{R}}\hat{S}_{\bm{R}}^x$ or $\sum_{\bm{R}}\hat{S}_{\bm{R}}^y$. 
Full SU(2) spin rotation symmetry is preserved only when both drives are linearly polarized and perfectly aligned, resulting in an isotropic effective spin model.
To further understand how each term in Eq.~\eqref{eq:XXZ} is optically induced, we refer to the processes illustrated in Fig.~\ref{Fig_1}(b). 
The interaction term $\sim \hat{S}_{\bm{R}}^z \hat{S}_{\bm{R}'}^z$, which splits singlet and triplet states without yielding spin exchange, arises from transitions between panels (i) and (ii) or between (ii) and (iii). 
Consequently, this interaction is always present unless the two drives are circularly polarized in opposite directions.  
In contrast, the flip-flop term $\hat{S}_{\bm{R}}^x\hat{S}_{\bm{R}'}^x+\hat{S}_{\bm{R}}^y\hat{S}_{\bm{R}'}^y$ originates from a series of processes that transforms from panel (i) to (iii). 
Therefore, this term requires both drives to contain components of both circular polarizations.

In the rest of this work, for the sake of concreteness we set $\Omega_{1,+} = \Omega_{1,-}$ and $\Omega_{2,+} =  e^{i\phi} \Omega_{2,-}$.
$\phi\in[-\pi,0]$ is the azimuthal angle defined on the Bloch sphere of the second drive, and can be used to control $\alpha = \cos(\phi)$.
Note that while the driven Zeeman splitting $B_L = 0$ in this parameter subspace, we will later introduce a real magnetic field $B$ to mimic the role of $B_L$.

\begin{figure}[t]
\centering
\includegraphics[width=\columnwidth]{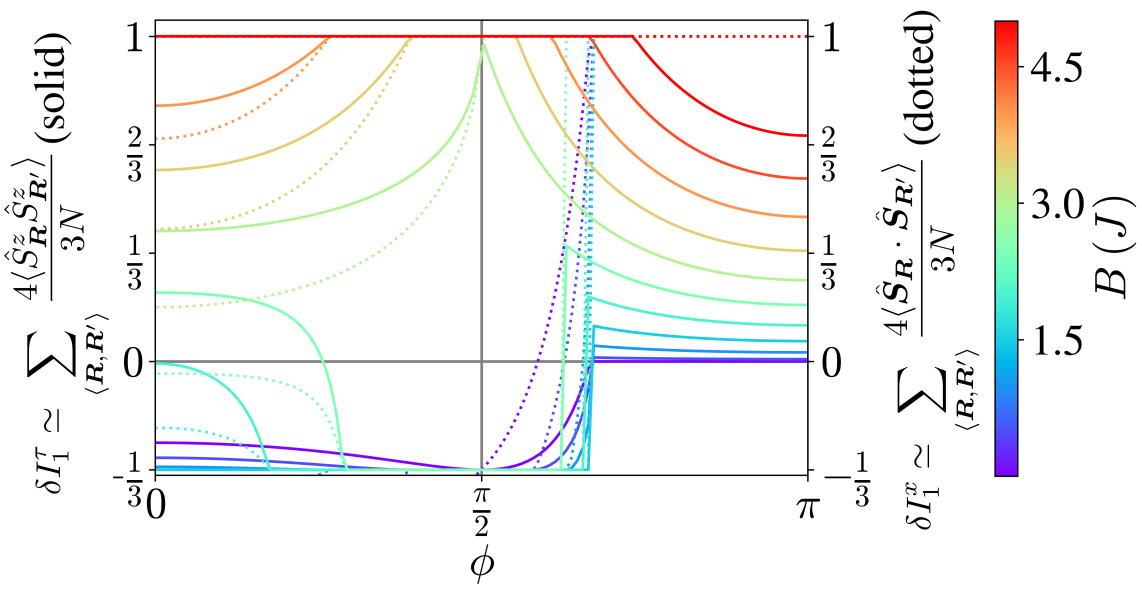}
\caption{Magnetic correlations from the classical mean-field ground state of Eq.~\eqref{eq:XXZ} (see End Matter) and the corresponding relative Stokes emission intensities for $\delta < 0$.
Colors label magnetic fields, with a minimum value $B = 10^{-5} J$ introduced to avoid the zero-field degeneracy.
}
\label{Fig_3}
\end{figure}

\textit{Optical detection.} --- 
The same setup can be utilized to measure spin correlations within the moiré system by detecting the fluctuations $\delta\hat{\bm{E}}(t)$.  
Although this approach generally applies to magnetism arising from various mechanisms (for instance, induced by intrinsic charge tunneling which we have neglected), we assume here that the drives are sufficiently strong such that Eq.~\eqref{eq:XXZ} dominates the magnetic correlations. 
In this context, it is crucial to be in compliance with the energy hierarchy given by Eq.~\eqref{eq:energy_hierarchy} to ensure the validity of the spin model.

The quantities of interest here are the intensities of the detected optical modes at frequency $\omega - U$ (also referred to as Stokes emission~\cite{gangopadhyay1992spectral}) with polarizations filtered along $\bm{e}_\tau$ and $\bm{e}_x$, denoted as $I_1^{\tau}$ and $I_1^{x}$.
Using standard input-output formalism~\cite{Supplement}, one can relate this intensity to the correlators $\frac{1}{N} \sum_{\langle\bm{R},\bm{R}'\rangle} \langle \hat{S}_{\bm{R}}^z \hat{S}_{\bm{R}'}^z \rangle$ and $\frac{1}{N} \sum_{\langle\bm{R},\bm{R}'\rangle} \langle \hat{\bm{S}}_{\bm{R}}\cdot\hat{\bm{S}}_{\bm{R}'}\rangle$, where $N$ is the number of moiré sites, $\hat{\bm{S}}_{\bm{R}} = (\hat{S}_{\bm{R}}^x, \hat{S}_{\bm{R}}^y, \hat{S}_{\bm{R}}^z)$, and the brackets denote the expectation value with respect to the driven state.
Their relative fraction of change $\delta I_1^{\mathrm{p}} \equiv 1 - \frac{I_1^{\mathrm{p}}}{\mathcal{I}_1^{\mathrm{p}}}$ ($\mathrm{p} \in \{x,\tau\}$), where $\mathcal{I}_1^{\mathrm{p}}$ is defined as the Stokes intensity $I_1^{\mathrm{p}}$ when the underlying spin configuration is paramagnetic, satisfy:
\begin{equation}
\label{eq:Stokes_measure}
\delta I_1^{\tau}
\simeq 
\sum_{\langle\bm{R},\bm{R}'\rangle} 
\frac{4\langle \hat{S}_{\bm{R}}^z \hat{S}_{\bm{R}'}^z \rangle}{3N} 
,\;
\delta I_1^{x}
\simeq 
\sum_{\langle\bm{R},\bm{R}'\rangle} 
\frac{4\langle \hat{\bm{S}}_{\bm{R}}\cdot\hat{\bm{S}}_{\bm{R}'}\rangle}{3N} 
.
\end{equation}
These relations can be qualitatively understood by noting that the effect of the detection photon on VMB electrons mimics that of $\Omega_{2,\tau}$ (except that it is incoherently emitted).
Therefore, our previous arguments for the form of the effective model Eq.~\eqref{eq:XXZ} using symmetries of light-matter coupling and Fig.~\ref{Fig_1}(b) applies similarly to Eq.~\eqref{eq:Stokes_measure}.
The corresponding measurement procedure is outlined as follows:
First, at the target temperature, measure the Stokes photon intensity with polarization along $\bm{e}_\tau$, yielding $I_1^{\tau}$. 
Then, raise the temperature to access a paramagnetic state and record the corresponding intensity, $\mathcal{I}_1^{\tau}$.
Next, repeat these measurements for Stokes photons polarized along $\bm{e}_x$, obtaining $I_1^{x}$ and $\mathcal{I}_1^{x}$. 
The magnetic correlators are then extracted using Eq.~\eqref{eq:Stokes_measure}.
Alternatively, spin correlators can be determined without information of paramagnetic state by measuring $I_1^{\mathrm{p}}$ in extra polarization modes $\bm{e}_{\mathrm{p}}$ (see Supplementary Material~\cite{Supplement}).

\textit{Signatures of the driven magnetic phases.} ---
A fingerprint of Eq.~\eqref{eq:XXZ} is the drastic change in its magnetic properties, and consequently in its optical responses, at the phase boundaries.
To gain some physical understanding on these sharp transitions, we apply the standard three-sublattice classical mean-field treatment to the $XXZ$ Hamiltonian, which we refer to End Matter for details.
Fig.~\ref{Fig_3} shows how the relative Stokes intensities, along with spin correlators, vary with $\phi$ at different magnetic fields $B$. 
Significant changes and non-differentiable points appear in these optical responses, marking the phase boundaries in Fig.~\ref{Fig_3}. 
In particular, sharp jumps in $\delta I_1^\tau$ and $\delta I_1^x$ occur within $B\in(1.5,3)J$, resulting from transitions between phases with opposite signs of $\frac{1}{N} \sum_{\langle\bm{R},\bm{R}'\rangle} \langle \hat{S}_{\bm{R}}^z \hat{S}_{\bm{R}'}^z \rangle$.
These abrupt changes serve as clear signatures of the driven $XXZ$ dynamics at low temperatures, and the discussion for finite temperature is left to the End Matter.

\textit{Discussion and Outlook.} --- 
Although our analysis focuses on nearest-neighbor tunneling and spin-spin interactions, a valid treatment for large superlattice spacing, long-range optically driven contributions may become relevant in bilayers with smaller moiré periods. 
Notably, including next-nearest-neighbor spin-spin interactions in triangular lattices could lead to quantum spin liquids~\cite{hu2015competing,saadatmand2015phase,zhu2015spin}. 
While similar predictions exist for undriven moiré bilayers~\cite{wu2018hubbard}, optical drives might enhance spin-spin interactions, boosting the stability of spin liquid states. Additionally, extending our two-photon probe method with time resolution could enable detection of these states~\cite{nambiar2024diagnosing}. 
More broadly, this technique might also aid studies of other spin liquid systems beyond TMDs~\cite{savary2016quantum}, such as those described by the Kitaev honeycomb model~\cite{kitaev2006anyons,claassen2017dynamical}.

Our scenario can be generalized to explore electronic states and dynamics away from half-filling.
For instance, in lightly hole-doped regimes, valley-dependent tunneling from Eq.\eqref{eq:t_operator} could induce spin-valley currents, beneficial for spintronics or valleytronics applications~\cite{schaibley2016valleytronics}.
Near half-filling, density-dependent tunneling enables independent tuning of charge excitation (doublons and holons~\cite{huang2023spin}) hoppings and spin-spin interactions.
This makes optically driven moiré TMDs promising for investigating phase diagrams of the paradigmatic t-J model~\cite{lee2006doping,auerbach2012interacting} and anisotropic variants like the t-J$_z$ model~\cite{dagotto1994correlated}.

An important feature of these models is the charge-spin interplay~\cite{schlomer2024kinetic,huang2023spin,huang2023mott}, which may be optically tuned. 
In particular, geometric frustration in triangular lattices makes magnetism sensitive to the sign of charge tunneling~\cite{ciorciaro2023kinetic,schlomer2024kinetic}, which in our approach is controlled by $\Delta$.
A similar effect may arise from dynamics of optical excitations (excitons)~\cite{wang2022light,ciorciaro2023kinetic,huang2023mott}, with the key distinction that their dipolar interaction-driven tunneling~\cite{shahmoon2017cooperative,moreno2021quantum,sierra2022dicke,pedersen2024green,kumlin2025superradiance} includes a valley-flip sector~\cite{huang2024collective}. 
While suppressed in the detuned regime, this effect is expected to grow as $\Delta$ approaches the exciton binding energy.
The kinetic magnetism from charge or exciton dynamics in optically driven TMDs remains an open question for future study.

Another promising direction stemming from our work is the potential application of our optical methods to manipulate other correlated states in moiré TMDs, such as Kondo insulators~\cite{zhao2023gate,guerci2023chiral,devakul2022quantum}, fractional Chern states~\cite{cai2023signatures,zeng2023thermodynamic,redekop2024direct}, and superconductivity~\cite{guo2024superconductivity,xia2024superconductivity}.
Finally, the optical approach itself can be further generalized by employing optical drives with spatial~\cite{andreoli2024metalens,sarkar2024sub,session2024optical} or temporal modulation~\cite{hung2016quantum}.
These possibilities greatly expand the scope for future exploration in strongly correlated electron-photon systems.

\textit{Acknowledgements.} --- 
We thanks Atac Imamoglu and Eugene Demler for useful discussions.

\bibliography{Biblio}

\clearpage

\appendix

\onecolumngrid

\section{End Matter}

\twocolumngrid

\textit{Appendix A: Classical mean-field theory for the $XXZ$ model.} ---
In this mean-field analysis, all sites $\{\bm{R}\}$ in the triangular lattice with periodicity $a_M$ are categoried into three sublattice sets, with each of them being a triangular lattice with enlarged periodicity $\sqrt{3}a_M$.
In addition, the operator $\hat{\bm{S}}_{\bm{R}}$ is treated as a classical vector characterized by its polar and azimuthal angles.
Therefore, the mean-field Hamiltonian contains six angular parameters. 
Notably, one of the three azimuthal angles can be gauged away, and therefore, effective there are only five parameters, and the mean-field ground state is determined by minimizing the energy with respect to them.
Below we summarize the results for the model $\hat{H}_{S} = - B \sum_{\bm{R}} \hat{S}_{\bm{R}}^z + \hat{H}_{J}$, with $\hat{H}_{J}$ given by Eq.~\eqref{eq:XXZ}.

We first summarize the solutions for $J>0$ (and $B>0$) following Refs.~\cite{miyashita1986magnetic,murthy1997superfluids,yamamoto2014quantum}.
In this case, the ground state solutions always give coplanar spins, and therefore, without the loss of generality, we can label these solutions with the polar angles (within $\pm\pi$) of the spins in the three sublattices, which we denote as $\vartheta_A$, $\vartheta_B$, and $\vartheta_C$.

For $0<\alpha<1$, the magnetic phases are separated by three critical magnetic fields $B_3>B_2>B_1>0$, where:
\begin{equation}
\begin{aligned}
B_1
&=
\frac{3J\alpha}{2}
\\
B_2
&=
\frac{3J}{2}
\left[
1-\frac{\alpha}{2}
+
\sqrt{
1+\alpha
-
\frac{7}{4}
\alpha^2
}
\right]
\\
B_3
&=
\frac{3J}{2}
(2+\alpha)
.
\end{aligned}
\end{equation}
The angular parameters from the mean-field solutions correspond to:
\begin{equation}
(\vartheta_A, \vartheta_B, \vartheta_C)
=
\begin{cases}
(\pi,\vartheta,-\vartheta)
&,\quad
0<B<B_1
\\
(\pi,0,0)
&,\quad
B_1<B<B_2
\\
(\vartheta',\vartheta'',\vartheta'')
&,\quad
B_2<B<B_3
\\
(0,0,0)
&,\quad
B>B_3
\end{cases}
,
\end{equation}
where $\vartheta = \cos^{-1}\left[\frac{1+\frac{2B}{3J}}{1+\alpha}\right]$, and $\vartheta'$ and $\vartheta''$ are determined by minimizing the following energy function:
\begin{equation}
\begin{aligned}
&E(\vartheta', \vartheta'')
=
\frac{J}{4}
[
(2\cos\vartheta'+\cos\vartheta'')\cos\vartheta''
\\
+ 
\alpha(2&\sin\vartheta'+\sin\vartheta'')\sin\vartheta''
] 
- 
\frac{B}{6}[\cos\vartheta' + 2\cos\vartheta'']
.
\end{aligned}
\end{equation}

Next, for $-0.5<\alpha<0$, there again exists three critical magnetic fields $\tilde{B}_3>\tilde{B}_2>\tilde{B}_1>0$, where:
\begin{equation}
\begin{aligned}
\tilde{B}_1
&=
\frac{3J}{4}
(2-\alpha-\sqrt{4+4\alpha-7\alpha^2})
\\
\tilde{B}_2
&=
J
[
1-\alpha+2\sqrt{(1-\alpha)(1+2\alpha)}
]
\\
\tilde{B}_3
&=
3J(1-\alpha)
,
\end{aligned}
\end{equation}
and the mean-field solutions to the ground state gives:
\begin{equation}
(\vartheta_A, \vartheta_B, \vartheta_C)
=
\begin{cases}
(\vartheta',\vartheta'',\vartheta'')
&,\quad
0<B<B_1
\\
(\pi,0,0)
&,\quad
B_1<B<B_2
\\
(\tilde{\vartheta},\tilde{\vartheta},\tilde{\vartheta})
&,\quad
B_2<B<B_3
\\
(0,0,0)
&,\quad
B>B_3
\end{cases}
,
\end{equation}
where $\tilde{\vartheta} = \cos^{-1}(\frac{B}{B_3})$.

Finally, for $-1<\alpha<-0.5$, there is only one critical magnetic field $B_3$, and the mean-field ground state corresponds to:
\begin{equation}
(\vartheta_A, \vartheta_B, \vartheta_C)
=
\begin{cases}
(\tilde{\vartheta},\tilde{\vartheta},\tilde{\vartheta})
&,\quad
0<B<B_3
\\
(0,0,0)
&,\quad
B>B_3
\end{cases}
.
\end{equation}

\begin{figure}[t]
\centering
\includegraphics[width=\columnwidth]{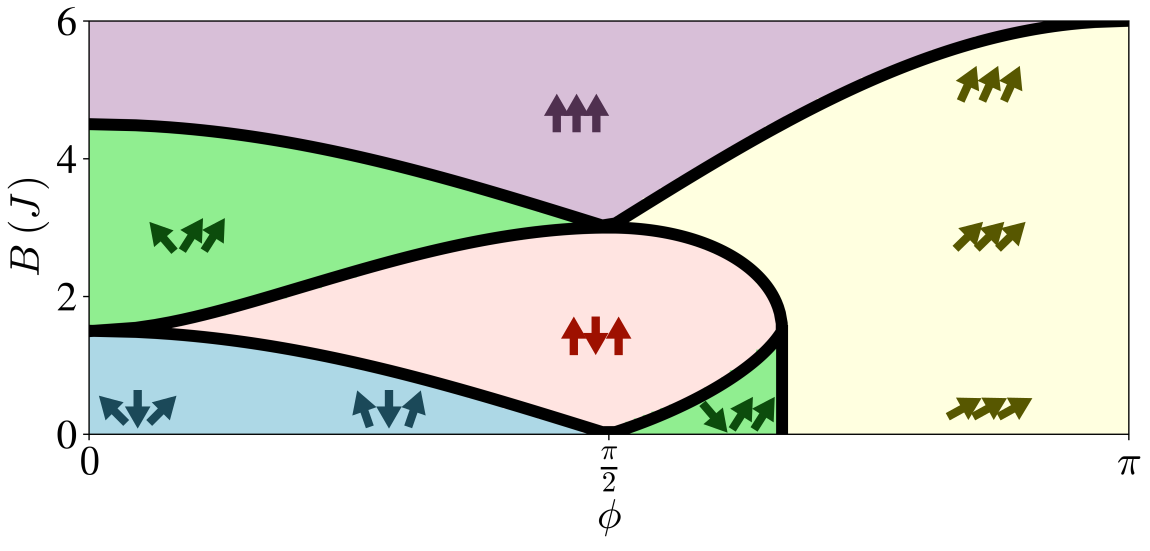}
\caption{Magnetic phase diagram from the classical three-sublattice mean-field ground state of Eq.~\eqref{eq:XXZ} for $J>0$, ranged over spin-spin interaction anisotropy $\alpha = \cos(\phi)$ and magnetic field $B$.
}
\label{Fig_4}
\end{figure}

The magnetic phase diagram from the above expressions (for $J>0$) is summarized in Fig.~\ref{Fig_4}.
It captures the essential qualitative features of the quantum phase diagram as obtained from numerical computations.
Specifically, the modifications after incorporating quantum fluctuations are (1) the phase boundary at $\alpha = -0.5$ becomes $B$-dependent, and (2) the critical point at $(\alpha, B) = (1, 1.5 J)$ split into two phase boundaries~\cite{yamamoto2014quantum}.
The case for $J < 0$ can be similarly analyzed, and its classical mean-field ground state simply gives a polarized spin states for all $|\alpha| \leq 1$.

\begin{figure}[h]
\centering
\includegraphics[width=\columnwidth]{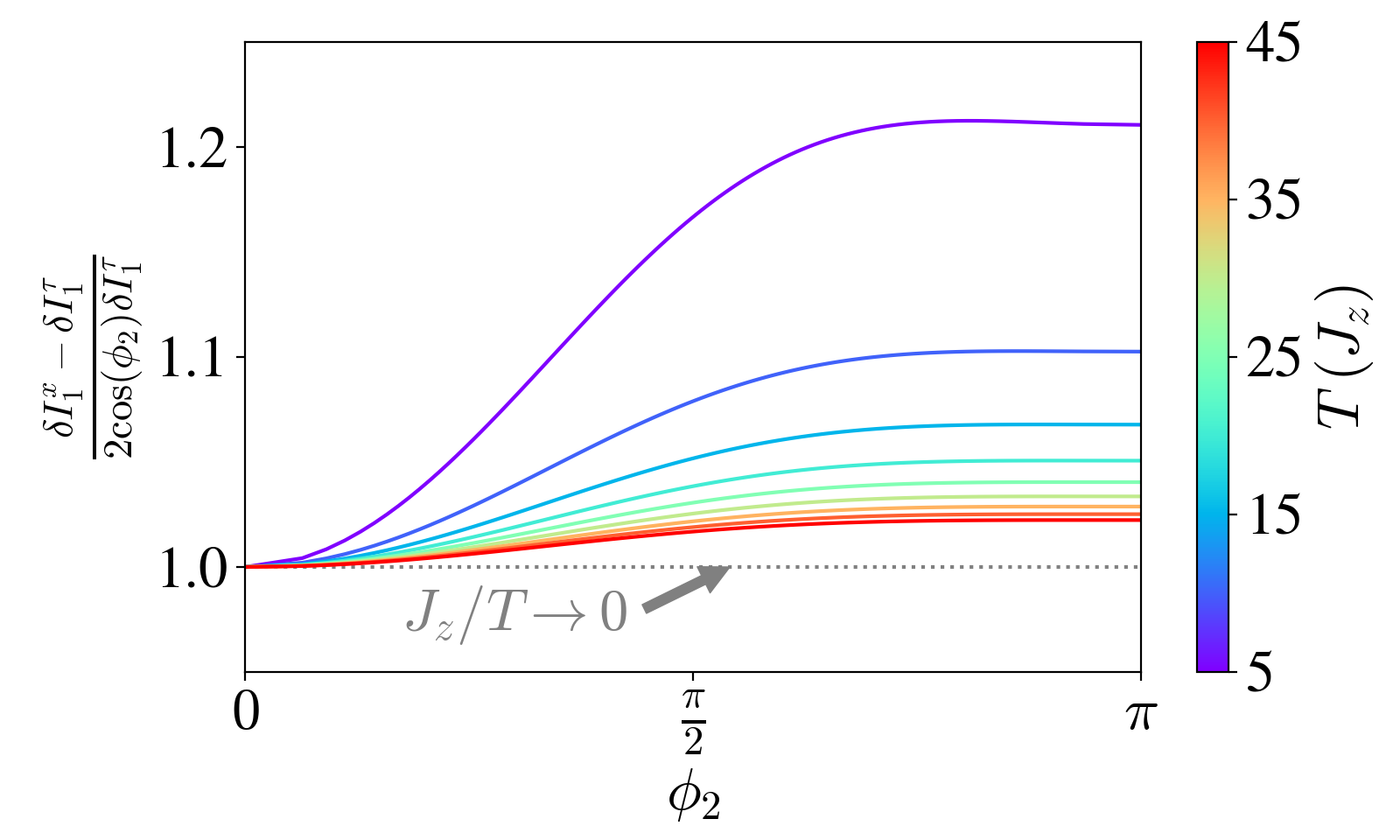}
\caption{
Ratio between the relative Stokes signals at finite temperature $T$ and zero magnetic field.
The magnetic correlators are evaluated with a thermal state expanded in high temperature series, up to subleading order in $J/T$.
$\Omega_{1,+} = \Omega_{1,-}$ and $\Omega_{2,+} =  e^{i\phi_2} \Omega_{2,-}$ is set in this figure.
Details are provided in the Supplementary Material~\cite{Supplement}.
}
\label{Fig_5}
\end{figure}

\textit{Appendix B: Finite temperature.} ---
In practice, limitations on the optical parameters might impose an upper bound on $J$, and the temperature $T$ could exceed this energy scale, which is beyond the scope of the mean-field descriptions Fig.~\ref{Fig_3} and Fig.~\ref{Fig_4}. 
To address this, we propose an alternative signature of $XXZ$ physics that can be observed at high temperatures, where $T \gg J$.
For simplicity, we focus on the case of $B = 0$.
In this high-temperature regime, although the phase diagram in Fig.~\ref{Fig_4} no longer applies, the ratio between the two spin correlators in Eq.~\eqref{eq:Stokes_measure}, and consequently $\frac{\delta I_1^x}{\delta I_1^\tau}$, can still serve as an indication of spin anisotropy.
Specifically, $\frac{\delta I_1^x}{\delta I_1^\tau}$ displays a nontrivial dependence on $\phi$ and approaches $1+2\alpha$ as $T$ increases, as demonstrated in Fig.~\ref{Fig_5}.
These features provide a means of benchmarking whether the driven $XXZ$ dynamics are accurately realized in real experimental setups.

\end{document}


\begin{CJK}{UTF8}{gbsn}

\title{Supplementary Material: Optical engineering and detection of magnetism in moir\'e semiconductors}

\author{Tsung-Sheng Huang}
\affiliation{Joint Quantum Institute, University of Maryland, College Park, MD 20742, USA}

\author{Andrey Grankin}
\affiliation{Joint Quantum Institute, University of Maryland, College Park, MD 20742, USA}

\author{Yu-Xin Wang (王语馨)}
\affiliation{Joint Center for Quantum Information and Computer Science, University of Maryland, College Park, MD 20742, USA}

\author{Mohammad Hafezi}
\affiliation{Joint Quantum Institute, University of Maryland, College Park, MD 20742, USA}
\affiliation{Joint Center for Quantum Information and Computer Science, University of Maryland, College Park, MD 20742, USA}

\date{\today}

\maketitle
\end{CJK}

\section{Derivation of the Light-matter coupling}
\label{sec:light-matter}

In this section, we derive the light-matter coupling in terms of superlattice degrees of freedom starting from the following microscopic description:
\begin{equation}
\hat{V}_{\mathrm{LM}}
=
-
\int
d^3\bm{r}
\hat{\bm{E}}(\bm{r})
\cdot
\hat{\bm{P}}(\bm{r})
,\quad
\hat{\bm{P}}(\bm{r})
=
-e
\bm{r}
\hat{\psi}^\dagger(\bm{r})
\hat{\psi}(\bm{r})
,
\end{equation}
where $\hat{\psi}(\bm{r})$ is the electron field operator evaluated at (three-dimensional) position $\bm{r}$ with electric charge $-e$.
$\hat{\bm{P}}(\bm{r})$ and $\hat{\bm{E}}(\bm{r})$ are the polarization and electric field operators, respectively.
In the following, we assume that the moir\'e period $a_M$ is much larger than the monolayer lattice spacing, and the electric field varies at a length scale much greater than both of them.
This allows us to address these scales separately.

To tackle the atomic scale, we decompose the electron field operator in terms of Bloch states of two (decoupled) monolayers, which are respectively labeled by their orbitals, valley pseudospins (denoted with $\tau\in\pm$), spins, and Bloch momenta.
We aim at the low energy regime where spin and valley indices are locked due to spin-orbit coupling in TMDs and only the lowest conduction and the highest valence orbitals of each layer are relevant (denoted as $c$ and $v$ respectively).
In addition, we assume that all relevant charges are \textit{intralayer} such that the layer label can be omitted.
With these considerations, we express:
\begin{equation}
\label{eq:electron_field}
\hat{\psi}(\bm{r})
\simeq
\sum_{\tau}
\sum_{\bm{p}_c}
\psi_{\tau\bm{K}+\bm{p}_c}^{c}(\bm{r})
\hat{\Psi}_{\tau\bm{K}+\bm{p}_c}^c
+
\sum_{\bm{p}_v}
\psi_{\tau\bm{K}+\bm{p}_v}^{v}(\bm{r})
\hat{\Psi}_{\tau\bm{K}+\bm{p}_v}^v
,
\end{equation}
where $\hat{\Psi}_{\tau\bm{K}+\bm{p}_c}^c$ and $\hat{\Psi}_{\tau\bm{K}+\bm{p}_v}^v$ are the conduction and valence electron annihilation operators respectively, with their corresponding electronic Bloch wavefunctions being $\psi_{\tau\bm{K}+\bm{p}_c}^{c}(\bm{r})$ and $\psi_{\tau\bm{K}+\bm{p}_v}^{v}(\bm{r})$.
We focus on the momenta near the Brillouin zone corners $\tau\bm{K}$ with $\bm{p}_n$ being the momentum relative to them.
Note that both charges share the same set of valley momenta due to the intralayer assumption.
Transforming these monolayer band degrees of freedom into the Wannier basis and then approximating atomic (in-plane) positions, denoted as $\bm{s}$, as continuous variables, we find:
\begin{equation}
\hat{V}_{\mathrm{LM}}
\simeq
-
\int
d^2\bm{s}
\hat{\bm{E}}(\bm{s})
\cdot
\hat{\bm{P}}(\bm{s})
,\quad
\hat{\bm{P}}(\bm{s})
=
\sum_{\tau}
\left[
\bm{d}_\tau
\hat{\Psi}_\tau^{c\dagger}(\bm{s})
\hat{\Psi}_\tau^{v}(\bm{s})
+
\mathrm{H.c.}
\right]
,\quad
\bm{d}_\tau
=
-e
\int
d^3\bm{r}
\psi_{\tau\bm{K}}^{c\ast}(\bm{r})
\bm{r}
\psi_{\tau\bm{K}}^{v}(\bm{r})
=
d
\bm{e}_\tau^\ast
,
\end{equation}
where $\hat{\Psi}_{\tau}^c(\bm{s})$ and $\hat{\Psi}_{\tau}^{v}(\bm{s})$ are Fourier transform of $\hat{\Psi}_{\tau\bm{K}+\bm{p}_c}^c$ and $\hat{\Psi}_{\tau\bm{K}+\bm{p}_v}^{v}$, respectively.
$\bm{d}_\tau$ is the transition dipole operator that we treated within the $\bm{k}\cdot\bm{p}$ approximation to leading order.
While finding its magnitude $d$ relies on more involved numerical calculation, its direction always aligns with $\bm{e}_\tau^\ast = \frac{1}{\sqrt{2}}(\bm{e}_x-i\tau\bm{e}_y)$, where $\bm{e}_x$ and $\bm{e}_y$ are Cartesian in-plane unit vectors, due to rotational symmetries of the wavefunctions.

We proceed by introducing moir\'e length scale into the above description.
To do this, we expand $\hat{\Psi}_\tau^{c}(\bm{s})$ and $\hat{\Psi}_\tau^{v}(\bm{s})$ into \textit{moir\'e}-Wannier orbitals (note that this is different from the monolayer Wannier orbitals).
We further focus on the regime where only the first conduction and valence moir\'e bands (labeled as CMB and VMB respectively) are relevant~\footnote{This assumption is valid when Coulomb attraction between electrons and holes are weaker than moir\'e potential.}, indicating: 
\begin{equation}
\hat{\Psi}_\tau^c(\bm{s})
=
\frac{1}{\sqrt{N}}
\sum_{\bm{L}}
W^c(\bm{s}-\bm{L})
\hat{c}_{\bm{L},\tau}
,\quad
\hat{\Psi}_\tau^v(\bm{s})
=
\frac{1}{\sqrt{N}}
\sum_{\bm{R}}
W^v(\bm{s}-\bm{R})
\hat{v}_{\bm{R},\tau}
,
\end{equation}
where $\bm{L}$ and $\bm{R}$ are triangular supersites for CMB electrons and VMB holes, respectively.
The two superlattices are generally different set of positions, although we assume their total number $N$ are the same for simplicity.
$W^c(\bm{s}-\bm{L})$ and $W^v(\bm{s}-\bm{R})$ are the corresponding moir\'e-Wannier orbitals, and $\hat{c}_{\bm{L},\tau}$ and $\hat{v}_{\bm{v},\tau}$ are the annihilation operators for these degrees of freedom, respectively.
The above expression yields:
\begin{equation}
\label{eq:V_LM}
\hat{V}_{\mathrm{LM}}
\simeq
-
\sum_\tau
\sum_{\bm{L},\bm{R}}
\hat{\bm{E}}\left(\frac{\bm{L}+\bm{R}}{2}\right)
\cdot
\left[
d
\bm{e}_\tau^\ast
\eta_{\bm{L},\bm{R}}
\hat{c}_{\bm{L},\tau}^\dagger
\hat{v}_{\bm{R},\tau}
+
\mathrm{H.c.}
\right]
,\quad
\eta_{\bm{L},\bm{R}}
=
\int
d^2\bm{s}
W^{c\ast}(\bm{s}-\bm{L})
W^v(\bm{s}-\bm{R})
.
\end{equation}
Note that the parameter $\eta_{\bm{L},\bm{R}}$ is expected to scale exponentially with $|\bm{L}-\bm{R}|$ and does not depend on $\bm{L}+\bm{R}$.
Eq.~\eqref{eq:V_LM} becomes Eq.~(3) in the main text after incorporating the two external drives in the interaction picture.
We utilize this expression to derive effective models and investigate measurement schemes.

\subsection{Wannier orbital overlapping}
\label{sec:Wannier_orbital_overlapping}

\begin{figure}[t]
\centering
\includegraphics[width=\columnwidth]{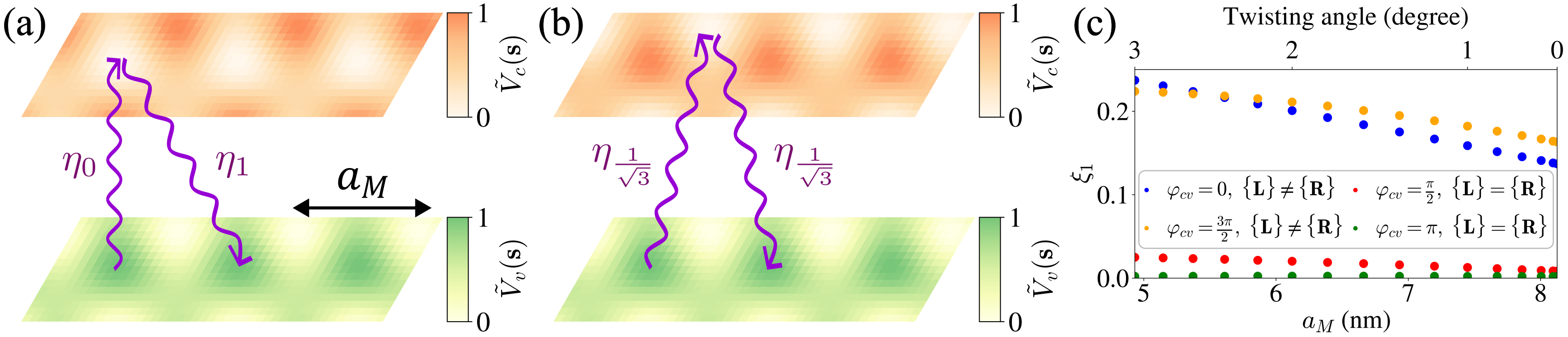}
\caption{
Illustration of (a) identical and (b) shifted lattices. 
Here $\tilde{V}_c(\bm{s})$ and $\tilde{V}_v(\bm{s})$ denote the normalized moir\'e potentials from $V_c(\bm{s})$ and $V_v(\bm{s})$, respectively.
Panel (c) shows $\xi_1$ at various $\varphi_{cv}$ and $a_M$.
Four representative values of $\varphi_{cv}$ are chosen to illustrate two identical and two shifted lattices.
Mor\'e period is related to twisting angle $\theta$ as $a_M(\theta) = \frac{a_{\mathrm{ML}}}{\sqrt{\theta^2+\delta_{\mathrm{ML}}^2}}$, where $a_{\mathrm{ML}}$ is the average of lattice constants from the two TMD layers and $\delta$ is the mismatch between them.
These parameters and others are set by the values for MoSe\textsubscript{2}/WS\textsubscript{2}:
$a_{\mathrm{ML}} = 0.325$nm and $\delta_{\mathrm{ML}} = 0.04$~\cite{kang2013band}.
The electron masses are $(m_c, m_v) = (0.43, -0.5)m_0$~\cite{conti2020transition} (with $m_0$ being the bare electron mass).
The potential parameters are $V = 7$meV and $\varphi_v = 35^\circ$~\cite{zhang2020moire}.
}
\label{Fig_S1}
\end{figure}

In this section, we provide estimations for the moir\'e-Wannier orbital overlapping $\eta_{\bm{L},\bm{R}}$ defined in Eq.~\eqref{eq:V_LM} as well as the quantity:
\begin{equation}
\label{eq:xi_1}
\xi_1 \equiv \xi_{\bm{R},\bm{R}'}|_{|\bm{R}-\bm{R}'|=a_M}
,\quad
\xi_{\bm{R},\bm{R}'}=\sum_{\bm{L}}
\eta_{\bm{L},\bm{R}}^\ast
\eta_{\bm{L},\bm{R}'}
,
\end{equation}
that eventually appears in the nearest-neighbor spin-spin interactions.
We start by numerically solving the Bloch functions from the following Schr\"{o}dinger equations capturing the motion of charges within a moir\'e potential ($\hbar = 1$ is set hereafter):
\begin{equation}
\left[
-
\frac{(\nabla_{\bm{s}}+i\bm{k})^2}{2m_c}
+
V_c(\bm{s})
-
\mathcal{E}_{\bm{k}}^c
\right]
u_{\bm{k}}^c(\bm{s})
=
\left[
-
\frac{(\nabla_{\bm{s}}+i\bm{k})^2}{2m_v}
+
V_v(\bm{s})
-
\mathcal{E}_{\bm{k}}^v
\right]
u_{\bm{k}}^v(\bm{s})
=
0
.
\end{equation}
Here $\bm{k}$ denotes the moir\'e-Bloch momentum.
$m_c > 0$ and $m_v < 0$ are the effective electronic masses in CMB and VMB, respectively, and $V_c(\bm{s})$ and $V_v(\bm{s})$ are the corresponding moir\'e potentials.
These potentials, as well as the electronic eigenfunctions $u_{\bm{k}}^c(\bm{s})$ and $u_{\bm{k}}^v(\bm{s})$, are invariant under a shift in $\bm{s}$ by superlattice vectors.
$\mathcal{E}_{\bm{k}}^c$ and $\mathcal{E}_{\bm{k}}^v$ represent the eigenvalues. 
Note that here we focus only on the lowest state for CMB and highest state for VMB for each $\bm{k}$, and therefore, the band indices are suppressed.

To proceed, we phenomenologically capture the moir\'e potentials for \textit{electrons} with the following expressions:~\cite{zhang2020moire}
\begin{equation}
V_c(\bm{s})
=
V
\sum_{i=1}^3
\cos\left(
\bm{G}_i\cdot\bm{s}+\varphi_c
\right)
,\quad
V_v(\bm{s})
=
V
\sum_{i=1}^3
\cos\left(
\bm{G}_i\cdot\bm{s}+\varphi_v
\right)
.
\end{equation}
Here $\bm{G}_1 = \frac{4\pi}{\sqrt{3}a_M}\bm{e}_y$, $\bm{G}_2 = \frac{2\pi}{a_M}(\bm{e}_x - \frac{\bm{e}_y}{\sqrt{3}})$, and $\bm{G}_3 = -\frac{2\pi}{a_M}(\bm{e}_x + \frac{\bm{e}_y}{\sqrt{3}})$.
$V$ denotes captures the strength of these potentials, which we assume to be the same for the two moir\'e bands, whereas their profiles are characterized by the phase factors $\varphi_c$ and $\varphi_v$.
Notably, whether the supersites for CMB and VMB ($\{\bm{L}\}$ and $\{\bm{R}\}$ respectively) are the same or different set of triangular lattice positions, which we refer to as \textit{identical lattices} and \textit{shifted lattices} scenarios, depends on the choices of these phases.
The situations $\varphi_{cv}\equiv\varphi_c - \varphi_v = \pi$ and $0$, yielding identical and shifted lattices respectively, are illustrated in Fig.~\ref{Fig_S1}(a) and (b), respectively.
In real experiments, each type of lattice configuration could potentially be achieved by accessing different stacking configurations~\cite{upadhyay2024giant,wang2023intercell}.

After solving the Bloch functions using the above potentials, the wavefunction overlap can be obtained via:
\begin{equation}
\eta_{\bm{L},\bm{R}}
=
\frac{1}{N}
\sum_{\bm{k}}
e^{i\bm{k}\cdot(\bm{L}-\bm{R})}
\eta_{\bm{k}}
,\quad
\eta_{\bm{k}}
\equiv
\int d^2\bm{s}
\,
u_{\bm{k}}^{c\ast}(\bm{s})
u_{\bm{k}}^{v}(\bm{s})
.
\end{equation}
It is worth noting that the phase of $\eta_{\bm{k}}$ cannot be determined solely by the Schr\"{o}dinger equations~\cite{marzari2012maximally}.
Nevertheless, different gauge choices generally lead to Wannier wavefunctions with different degree of localization; therefore, to validate the tight-binding approximation where $\eta_{\bm{L},\bm{R}}$ with large $|\bm{L}-\bm{R}|$ is negligible, one has to fix the phase to ensure localized Wannier wavefunctions.
Here we employ a simple approach to fix the phase of $u_{\bm{k}}^{v}(\bm{s})$, and similarly $u_{\bm{k}}^{c}(\bm{s})$, by that of $\int d^2\bm{s}\, u_{\bm{k}}^{v\ast}(\bm{s})f(\bm{s})$, with $f(\bm{s})$ some trial functions which we choose to be a real Gaussian.

Within this treatment, we numerically confirm that $\eta_{\bm{L},\bm{R}}$ overall decrease with larger $|\bm{L}-\bm{R}|$.
Importantly, this allows us to truncate the $\bm{L}$ summation in Eq.~\eqref{eq:xi_1} such that $\xi_1$ can be systematically computed.
By taking only the leading contribution, we find $\xi_1 = \eta_1^\ast\eta_0 +\eta_0^\ast\eta_1$ for identical lattice with $\eta_0$ and $\eta_1$ being $\eta_{\bm{L},\bm{R}}$ at $|\bm{L}-\bm{R}| = 0$ and $a_M$, respectively.
In contrast, $\xi_1 = \left|\eta_{\frac{1}{\sqrt{3}}}\right|^2$ for shifted lattice with $\eta_{\frac{1}{\sqrt{3}}}$ being $\eta_{\bm{L},\bm{R}}$ at $|\bm{L}-\bm{R}| = \frac{a_M}{\sqrt{3}}$.

Fig.~\ref{Fig_S1}(c) illustrates the estimated values of $\xi_1$ using parameters from MoSe\textsubscript{2}/WS\textsubscript{2}. 
Specifically, we fix $m_c$, $m_v$, $V$, and $\varphi_v$ while treating $a_M$ and $\varphi_c$ as tuning parameters. 
The flexibility in $\varphi_c$ allows for the illustration of both identical ($\varphi_c = \frac{\pi}{2}$ and $\pi$) and shifted ($\varphi_c = 0$ and $\frac{3\pi}{2}$) lattices. 
Generally, shifted lattices yield larger $\xi_1$ than identical ones, with differences reaching orders of magnitude for certain parameter sets.  
Overall, $\xi_1$ decreases as $a_M$ increases, which can be qualitatively understood by the narrowing of Wannier function envelopes relative to $a_M$. Additionally, interference between Wannier function ripples can introduce non-monotonic fluctuations in $\xi_1(a_M)$, as seen in the case of $\varphi_{cv} = \pi$. 
Notably, for this setting, the Wannier functions $W^c(\bm{s}-\bm{L})$ and $W^v(\bm{s}-\bm{R})$ are nearly orthogonal when $\bm{L} \neq \bm{R}$ (and fully orthogonal if $m_c = -m_v$), leading to a significantly suppressed $\eta_1$ and consequently $\xi_1$.

\section{Measurement of the Raman-induced spin order}
\label{sec:measure}

Here we demonstrate how to extract the magnetic correlations of VMB electrons that is induced by the two classical light sources via optical measurements.
The steps of the theoretical derivation is discussed respectively in the following sections:
In Section~\ref{sec:input-output} we first establish the input-output formalism for light scattering from electrons experiencing Hubbard interaction.
We then project the scattered electrmagnetic fields into detected modes in Section~\ref{sec:detected_mode}, and construct the intensities of these modes in Section~\ref{sec:Intensity_Stokes}.
Finally, for comparison, we discuss a different measurement scheme targeting at the magnetic correlations that originates from the intrinsic charge motion (rather from the drives) in Section~\ref{sec:measure_intrinsic_spin_order}.

\subsection{Derivation of input-output relation}
\label{sec:input-output}

In this section, we derive the input-output relation following Refs.~\cite{gruner1996green,dung2002resonant,asenjo2017atom}.
We start with the total electric-field operator that is expressed in terms of a superposition of sectors at different frequencies $\omega$:
\begin{equation}
\hat{\bm{E}}(\bm{r})
=
\int d\omega
\hat{\bm{E}}(\bm{r},\omega)
,\quad
\hat{\bm{E}}(\bm{r},\omega)
=
\hat{\bm{E}}^+(\bm{r},\omega)
+
\hat{\bm{E}}^-(\bm{r},\omega)
,\quad
\hat{\bm{E}}^+(\bm{r},\omega)
=
[\hat{\bm{E}}^-(\bm{r},\omega)]^\dagger
,
\end{equation}
where the frequency components has the following representation:
\begin{equation}
\label{eq:E_+_omega}
\hat{\bm{E}}^+(\bm{r},\omega)
=
i\mu_0\omega^2
\sqrt{\frac{\epsilon_0}{\pi}}
\int d^3\bm{r}'
\sqrt{\mathrm{Im}\{\epsilon(\bm{r}',\omega)\}}
\bm{\mathcal{G}}^{\epsilon}(\omega;\bm{r}-\bm{r}')
\cdot
\hat{\bm{f}}(\bm{r}',\omega)
.
\end{equation}
Here $\epsilon_0$ and $\mu_0$ are electric permittivity and magnetic permeability of free space, respectively, and at this point we keep the dielectric function $\epsilon(\bm{r},\omega)$ to be general and be position and frequency dependent.
$\bm{\mathcal{G}}^{\epsilon}(\omega;\bm{r}-\bm{r}')$ is the Green's tensor (propagator of electromagnetic field) that depends on the dielectric environment, which we emphasize with the superscript $\epsilon$.
$\hat{\bm{f}}(\bm{r},\omega)$ are vectors of bosonic operators with Cartesian components $\{\hat{f}_\alpha(\bm{r},\omega)|\alpha\in(x,y,z)\}$ satisfying $[\hat{f}_\alpha(\bm{r},\omega), \hat{f}_{\alpha'}(\bm{r}',\omega')] = \delta_{\alpha,\alpha'} \delta^{(3)}(\bm{r}-\bm{r}') \delta(\omega-\omega')$.

The $\hat{\bm{f}}(\bm{r},\omega)$ operators can be related the matter degrees of freedom through the following Heisenberg equation of motion:
\begin{equation}
\label{eq:Heisenberg_eom_f}
\partial_t
\check{\bm{f}}(\bm{r},\omega,t)
=
-i 
[\check{\bm{f}}(\bm{r},\omega,t), \check{H}(t)]
,\quad
\check{O}(t)
=
e^{i\hat{H}t} 
\hat{O} 
e^{-i\hat{H}t}
,\quad
\hat{H}
=
\hat{H}_{\mathrm{M}}
+
\hat{H}_{\mathrm{ph}}
+
\hat{V}_{\mathrm{LM}}
.
\end{equation}
Here $\hat{H}_{\mathrm{ph}} = \int d^3\bm{r} \int_0^\infty d\omega \omega \hat{\bm{f}}^\dagger(\bm{r},\omega) \hat{\bm{f}}(\bm{r},\omega)$ is the Hamiltonian for free electromagnetic field and $\hat{H}_{\mathrm{M}}$ is the matter sector.
Solution to the above equation reads:
\begin{equation}
\label{eq:check_f}
\check{\bm{f}}(\bm{r},\omega,t)
=
\check{\bm{f}}_0(\bm{r},\omega,t)
+
\frac{\omega^2 d^\ast}{c^2}
\sqrt{\frac{\mathrm{Im}\{\epsilon(\bm{r},\omega)\}}{\pi\epsilon_0}}
\sum_{\tau,\bm{L},\bm{R}}
\eta_{\bm{L},\bm{R}}^\ast
\int_0^t 
dt'
e^{-i\omega(t-t')}
\check{v}_{\bm{R},\tau}^\dagger(t')
\check{c}_{\bm{L},\tau}(t')
\bm{e}_\tau
\cdot
\bm{\mathcal{G}}^{\epsilon\ast}\left(\omega;\frac{\bm{L}+\bm{R}}{2}-\bm{r}\right)
,
\end{equation}
where $\check{\bm{f}}_0(\bm{r},\omega,t)$ denotes the solution in free space.

We proceed by making assumptions regarding time scales to simplify the above integral.
Specifically, we will assume that the time evolution of matter operators is dominated by the following matter Hamiltonian at short time scales:
\begin{equation}
\hat{H}_{\mathrm{M}}
=
U
\sum_{\bm{R}}
\hat{n}_{\bm{R},+}
\hat{n}_{\bm{R},-}
+
E_g
\sum_{\bm{L},\tau}
\hat{c}_{\bm{L},\tau}^\dagger
\hat{c}_{\bm{L},\tau}
,\quad
\hat{n}_{\bm{R},\tau}
=
\hat{v}_{\bm{R},\tau}^\dagger
\hat{v}_{\bm{R},\tau}
,
\end{equation}
where $U$ is the Hubbard interaction in VMB and $E_g$ is the bandgap.
Note that here we drop the quartic interactions between CMB electrons as we assume they are dilute, and suppress the tunneling of charges for simplicity.
Typically $E_g\sim 1$eV and $U\sim10$meV.
As both $E_g$ and $E_g - U$ are much faster than the spontaneous emission rate $\gamma = \frac{|d|^2E_g^3}{3\pi\epsilon_0c^3}\ll 1$meV~\cite{huang2024collective}, with $c$ denoting the speed of light, the Markov approximation is valid, giving $\check{v}_{\bm{R},\tau}^\dagger(t') \check{c}_{\bm{L},\tau}(t') \simeq e^{-i(t'-t) (E_g - U \hat{n}_{\bm{R},-\tau})} \check{v}_{\bm{R},\tau}^\dagger(t) \check{c}_{\bm{L},\tau}(t)$~\footnote{External drives are detuned following Eq.~(4) of the main text such that they effectively only provide corrections to the energy of electron-hole pairs that are much smaller than $E_g$.}.
In addition, in the spirit of coarse-grained time-averaging, we replace:
\begin{equation}
\int_0^t 
dt'
e^{-i(\omega - E_g + U \hat{n}_{\bm{R},-\tau})(t-t')} 
\to
\zeta(E_g - U \hat{n}_{\bm{R},-\tau} - \omega)
,\quad
\zeta(x)
\equiv
\pi
\delta(x)
+
i
\hat{\mathcal{P}}
\left(
\frac{1}{x}
\right)
,
\end{equation}
where $\hat{\mathcal{P}}(1/x)$ stands for the principal value of $1/x$ upon integration.
Subtituting this to Eq.~\eqref{eq:check_f} and rotate back to the Schroedinger representation, we find:
\begin{equation}
\hat{\bm{f}}(\bm{r},\omega)
\simeq
\hat{\bm{f}}_0(\bm{r},\omega)
+
\frac{\omega^2 d^\ast}{c^2}
\sqrt{\frac{\mathrm{Im}\{\epsilon(\bm{r},\omega)\}}{\pi\epsilon_0}}
\sum_{\tau,\bm{L},\bm{R}}
\zeta(E_g - U \hat{n}_{\bm{R},-\tau} - \omega)
\eta_{\bm{L},\bm{R}}^\ast
\hat{v}_{\bm{R},\tau}^\dagger
\hat{c}_{\bm{L},\tau}
\bm{e}_\tau
\cdot
\bm{\mathcal{G}}^{\epsilon\ast}\left(\omega;\frac{\bm{L}+\bm{R}}{2}-\bm{r}\right)
.
\end{equation}
Further putting this back to Eq.~\eqref{eq:E_+_omega} and exploiting the following integral relation for components of the Green's tensor:
\begin{equation}
\frac{\omega^2}{c^2}
\int d^3\bm{r}'
\mathrm{Im}\{\epsilon(\bm{r}',\omega)\}
\mathcal{G}_{\alpha,\beta}^{\epsilon}(\omega;\bm{r}-\bm{r}')
\mathcal{G}_{\alpha',\beta}^{\epsilon\ast}(\omega;\bm{R}-\bm{r}')
=
\mathrm{Im}
\left[
\mathcal{G}_{\alpha,\alpha'}^{\epsilon}(\omega;\bm{r}-\bm{R})
\right]
,
\end{equation}
we find:
\begin{equation}
\hat{\bm{E}}^+(\bm{r},\omega)
=
\hat{\bm{E}}_0^+(\bm{r},\omega)
+
\frac{i\mu_0d^\ast\omega^2}{\pi}
\sum_{\tau,\bm{L},\bm{R}}
\eta_{\bm{L},\bm{R}}^\ast
\mathrm{Im}
\left[
\bm{\mathcal{G}}^{\epsilon}\left(\omega;\bm{r}-\frac{\bm{L}+\bm{R}}{2}\right)
\right]
\cdot
\bm{e}_\tau
\zeta(E_g - U \hat{n}_{\bm{R},-\tau} - \omega)
\hat{v}_{\bm{R},\tau}^\dagger
\hat{c}_{\bm{L},\tau}
,
\end{equation}
where $\hat{\bm{E}}_0^+(\bm{r},\omega)$ is the free field associated to $\hat{\bm{f}}_0(\bm{r},\omega)$.
The total electric field can be obtained by integrating the above expression with $\omega$.
To simplify this integral, we note that for any function $f(\omega + i\nu)$ that is analytic in the complex plane $(\omega,\nu)$, one has the (Kramers-Kronig) relation $\mathrm{Re}[f(\omega)] = \frac{1}{\pi} \int_{-\infty}^{\infty} d\omega' \hat{\mathcal{P}}\left( \frac{\mathrm{Im}[f(\omega')]}{\omega'-\omega} \right) $~\cite{steck2007quantum}, which eventually yields:
\begin{equation}
\hat{\bm{E}}^+(\bm{r})
=
\hat{\bm{E}}_0^+(\bm{r})
+
\mu_0d^\ast
\sum_{\tau,\bm{L},\bm{R}}
\eta_{\bm{L},\bm{R}}^\ast
(E_g - U \hat{n}_{\bm{R},-\tau})^2
\bm{\mathcal{G}}^{\epsilon}\left(E_g + U \hat{n}_{\bm{R},-\tau};\bm{r}-\frac{\bm{L}+\bm{R}}{2}\right)
\cdot
\bm{e}_\tau
\hat{v}_{\bm{R},\tau}^\dagger
\hat{c}_{\bm{L},\tau}
.
\end{equation}
Note that the second term contains non-linearity that comes from Hubbard interaction.
In the regime $U\ll\omega_{\mathrm{ex}}$, this simplifies to:
\begin{equation}
\label{eq:input_output}
\hat{\bm{E}}^+(\bm{r})
=
\hat{\bm{E}}_0^+(\bm{r})
+
\hat{\bm{E}}_{sc}^+(\bm{r})
,\quad
\hat{\bm{E}}_{sc}^+(\bm{r})
\simeq
\mu_0
d^\ast
E_g^2
\sum_{\tau,\bm{L},\bm{R}}
\eta_{\bm{L},\bm{R}}^\ast
\bm{\mathcal{G}}^{\epsilon}\left(E_g;\bm{r}-\frac{\bm{L}+\bm{R}}{2}\right)
\cdot
\bm{e}_\tau
\hat{v}_{\bm{R},\tau}^\dagger
\hat{c}_{\bm{L},\tau}
,
\end{equation}
where the expression of scattered field $\hat{\bm{E}}_{sc}^+(\bm{r})$ is reminiscent of the standard input-output relation for atomic arrays~\cite{manzoni2018optimization}.

\subsection{Detection mode theory}
\label{sec:detected_mode}

In real experiments, photon scattering is achieved using light beams with specific profiles (such as Gaussian), and only the same mode from the scattered fields is collected.
For this reason, we project Eq.~\eqref{eq:input_output} into such detected mode. 
The beam waist is assumed to be much larger than moir\'e period such that spatial variation of mode profile can be neglected, and the remaining nontrivial degree of freedom is its polarization.
More specifically, we consider filtering the light to polarization $\bm{e}_{\mathrm{p}}$ (subscript $\mathrm{p}$ stands for ``probed mode'') before it goes into the detector, which collects light at a plane parallel to the sample at out-of-plane coordinate $z=z_{\mathrm{p}}$.
The associated field operators are~\cite{manzoni2018optimization}:
\begin{equation}
\label{eq:E_sc_d}
\hat{E}_{\mathrm{p}}^+
\equiv
\sqrt{\frac{2c\epsilon_0}{E_gA_{\mathrm{p}}}}
\bm{e}_{\mathrm{p}}^\ast
\cdot
\int_{z=z_{\mathrm{p}}} 
d^2\bm{r}
\hat{\bm{E}}^+(\bm{r})
=
\hat{E}_{0,\mathrm{p}}^+
+
\hat{E}_{sc,\mathrm{p}}^+
,\quad
\hat{E}_{sc,\mathrm{p}}^+
=
i
d^\ast
\sqrt{\frac{E_g}{2c\epsilon_0 A_{\mathrm{p}}}}
e^{\frac{iE_g|z_{\mathrm{p}}|}{c}}
\sum_{\tau,\bm{L},\bm{R}}
(
\bm{e}_{\mathrm{p}}^\ast
\cdot
\bm{e}_\tau
)
\eta_{\bm{L},\bm{R}}^\ast
\hat{v}_{\bm{R},\tau}^\dagger
\hat{c}_{\bm{L},\tau}
,
\end{equation}
where $A_{\mathrm{p}}$ is the area in the $z_{\mathrm{p}}$ plane in which light is collected.
Note that a normalization factor $\sqrt{\frac{2c\epsilon_0}{E_gA_{\mathrm{p}}}}$ is incorporated such that $\hat{E}_{\mathrm{p}}^-\hat{E}_{\mathrm{p}}^+$ (again the operators with $+$ and $-$ superscripts are Hermitian conjugate to each other) has the notion of photon per unit time.
$\hat{E}_{0,\mathrm{p}}^+ \equiv \sqrt{\frac{2c\epsilon_0}{E_gA_{\mathrm{p}}}}\int_{z=z_\mathrm{p}} d^2\bm{r} \bm{e}_{\mathrm{p}}^\ast \cdot \hat{\bm{E}}_0^+(\bm{r}) = \sqrt{\frac{2cA_{\mathrm{p}}\epsilon_0}{E_g}} \bm{e}_{\mathrm{p}}^\ast \cdot \hat{\bm{E}}_0^+$ is the according projected homogeneous free field.
As frequency resolution is required (see later in Section~\ref{sec:Intensity_Stokes}), the target operator is:
\begin{equation}
\label{eq:E_sc_d_+_omega}
\check{E}_{sc,\mathrm{p}}^+(\omega)
\equiv
\frac{1}{T}
\int_{t_0}^{t_0+T}
dt
e^{i\omega t}
\check{E}_{sc,\mathrm{p}}^+(t)
,\quad
\check{O}(t) 
= 
e^{i\hat{H}t} \hat{O} e^{-i\hat{H}t}
,
\end{equation}
where $t_0$ is the time when the detector starts to collect data and $T$ is the time-interval for collecting data which we eventually consider as infinite.
$\hat{O}$ is some generic operator in the Schr\"{o}dinger picture.
To evaluate $\check{E}_{sc,\mathrm{p}}^+(t)$, we consider the following equation of motion:
\begin{equation}
\partial_t
[
\check{v}_{\bm{R},\tau}^\dagger(t) \check{c}_{\bm{L},\tau}(t)
]
=
(-i)
[E_g - U \check{n}_{\bm{R},-\tau}(t)]
\check{v}_{\bm{R},\tau}^\dagger(t) \check{c}_{\bm{L},\tau}(t)
+
id
\bm{e}_\tau^\ast
\cdot
\sum_{\bm{R}'}
\eta_{\bm{L},\bm{R}'}
\check{\bm{E}}^+\left(\frac{\bm{L}+\bm{R}'}{2}, t\right)
\check{v}_{\bm{R},\tau}^\dagger(t) \check{v}_{\bm{R}',\tau}(t)
,
\end{equation}
where we drop the counter-rotating contributions in the last term.
Substitution of Eq.~\eqref{eq:input_output} yields:
\begin{equation}
\label{eq:vc_eom_with_Kernel}
\left[
\partial_t
+
i
\check{h}_{\bm{R},\tau}(t)
\right]
\check{v}_{\bm{R},\tau}^\dagger(t) \check{c}_{\bm{L},\tau}(t)
+
\gamma
\sum_{\bm{L}',\bm{R}',\tau'}
\check{\mathcal{K}}_{\bm{L}',\bm{R}',\tau'}^{\bm{L},\bm{R},\tau}(t)
\check{v}_{\bm{R}',\tau'}^\dagger(t) \check{c}_{\bm{L}',\tau'}(t)
=
id
\sum_{\bm{R}'}
\eta_{\bm{L},\bm{R}'}
\bm{e}_\tau^\ast
\cdot
\check{\bm{E}}_0^+(t)
\check{v}_{\bm{R},\tau}^\dagger(t) \check{v}_{\bm{R}',\tau}(t)
,
\end{equation}
where we denote:
\begin{equation}
\hat{h}_{\bm{R},\tau}=E_g - U \hat{n}_{\bm{R},-\tau}-\frac{i\gamma}{2}\sum_{\bm{R}'\neq\bm{R}}|\eta_{\bm{L},\bm{R}'}|^2\hat{n}_{\bm{R}',\tau}
,\quad
\gamma = \frac{|d|^2E_g^3}{3\pi\epsilon_0c^3}
,
\end{equation}
and the dimensionless kernel $\check{\mathcal{K}}_{\bm{L}',\bm{R}',\tau'}^{\bm{L},\bm{R},\tau}(t)$, originating from the off-site sector of the Green's tensor, only depends on VMB Heisenberg operators.
To proceed, we consider only bare dynamics for $\check{h}_{\bm{R},\tau}(t)$, which essentially gives $\hat{h}_{\bm{R},\tau}$, and take the input-field as the two-frequency ($\omega_\lambda$ with $\lambda\in\{1,2\}$, both comparable to $E_g$) classical drives $\check{\bm{E}}_0^+(t)\to\frac{1}{d}\sum_{\lambda=1}^2\sum_{\tau = \pm} e^{-i\omega_\lambda t}\Omega_{\lambda,\tau}\bm{e}_\tau$ (with $\Omega_{\lambda,\tau}$ denoting the components of the drives).
Subsequently, we time-integrate Eq.~\eqref{eq:vc_eom_with_Kernel} within the assumption that the motion within VMB are considerably slower than interband dynamics, which scale as $e^{-iE_g t}$, to implement Born-Markov approximation.
Focusing on long-time dynamics (which can be achieved by setting $t_0$ in Eq.~\eqref{eq:E_sc_d_+_omega} to be far enough from initial time) and the regime where the Raman detuning $\Delta\equiv E_g - \omega_1$ satisfies $|\Delta|\gg U\gg\gamma$ (which suppresses the kernel), we find:
\begin{equation}
\check{v}_{\bm{R},\tau}^\dagger(t) \check{c}_{\bm{L},\tau}(t)
\simeq
\frac{d}{\Delta}
\bm{e}_\tau^\ast
\cdot
\check{\bm{E}}_0^+(t)
\sum_{\bm{R}'}
\eta_{\bm{L},\bm{R}'}
\check{v}_{\bm{R},\tau}^\dagger(t) \check{v}_{\bm{R}',\tau}(t)
.
\end{equation}
Plugging this expression back to Eq.~\eqref{eq:E_sc_d} and Eq.~\eqref{eq:E_sc_d_+_omega}, we obtain:
\begin{equation}
\label{eq:scattered_mode_w}
\check{E}_{sc,\mathrm{p}}^+(\omega)
=
\frac{id^\ast e^{\frac{iE_g|z_{\mathrm{p}}|}{c}}}{T\Delta}
\sqrt{\frac{E_g}{2\epsilon_0c A_{\mathrm{p}}}}
\sum_{\lambda=1}^2
\sum_{\tau}
(
\bm{e}_{\mathrm{p}}^\ast
\cdot
\bm{e}_\tau
)
\Omega_{\lambda,\tau}
\sum_{\bm{R},\bm{R}'}
\xi_{\bm{R},\bm{R}'}
\hat{v}_{\bm{R},\tau}^\dagger \hat{v}_{\bm{R}',\tau}
\delta(\omega - \omega_\lambda + U \hat{n}_{\bm{R},-\tau} - U \hat{n}_{\bm{R}',-\tau})
,
\end{equation}
which will be utilized to compute the frequency-resolved emission intensity in the following section.

\subsection{Intensities of Stokes scattering}
\label{sec:Intensity_Stokes}

The standard quantity measured in frequency-resolved optical experiments is the intensity of detected modes.
For simplicity, we assume that the scattered mode can be separated from the input one (for instance, using interference technique), and therefore, the observable of interest here is $I_{sc,\mathrm{p}}(\omega) = \langle \check{E}_{sc,\mathrm{p}}^-(\omega) \check{E}_{sc,\mathrm{p}}^+(\omega) \rangle$.
Importantly, the delta function in Eq.~\eqref{eq:scattered_mode_w} indicates that $I_{sc,\mathrm{p}}(\omega)$ is nonzero only when $\omega = \omega_\lambda$, $\omega_\lambda + U$, and $\omega_\lambda - U$, which resembles Rayleigh and Stokes scattering~\cite{steck2007quantum,gangopadhyay1992spectral}, respectively.
We will focus only on the latter signal as it is relevant to superexchange (discussed later).
These considerations yields $I_{sc,\mathrm{p}}(\omega) \simeq I_\lambda^{\mathrm{p}} \delta(\omega - \omega_\lambda + U)$, where the intensity $I_\lambda^{\mathrm{p}}$ has the following form if the doublons are dilute for the reference state:
\begin{equation}
I_\lambda^{\mathrm{p}}
\sim
\frac{1}{A_{\mathrm{p}}}
\sum_{\tau,\tau'}
\Phi_{\lambda,\tau}^{\mathrm{p}\ast}
\Phi_{\lambda,\tau'}^{\mathrm{p}}
\sum_{\bm{R}\neq\bm{R}'}
|\xi_{\bm{R},\bm{R}'}|^2
\langle
\hat{v}_{\bm{R}',\tau}^\dagger \hat{v}_{\bm{R},\tau}
\hat{n}_{\bm{R},-\tau}^v
\hat{n}_{\bm{R},-\tau'}^v
\hat{v}_{\bm{R},\tau'}^\dagger \hat{v}_{\bm{R}',\tau'}
\rangle
,
\end{equation}
where $\Phi_{\lambda,\tau}^{\mathrm{p}} \equiv(\bm{e}_{\mathrm{p}}^\ast \cdot \bm{e}_\tau)
\Omega_{\lambda,\tau}$ characterizes the projection of the drives to the detection mode.
Here we suppress the proportionality factor, which does not explicitly depend on the target state, the drives $\lambda$, and the parameters labeled by the detected mode; therefore it can be calibrated by measuring $I_\lambda^{\mathrm{p}}$ on another known spin configuration (see below).
Considering coherent state for photons and taking only nearest-neighbor terms, we find:
\begin{equation}
\label{eq:I_lambda_p}
I_\lambda^{\mathrm{p}}
\sim
\frac{1}{A_{\mathrm{p}}}
\sum_{\tau}
\left[
\frac{3N}{4}
|\Phi_{\lambda,\tau}^{\mathrm{p}}|^2
+
\sum_{\langle\bm{R},\bm{R}'\rangle} 
(
\Phi_{\lambda,\tau}^{\mathrm{p}\ast} 
\Phi_{\lambda,-\tau}^{\mathrm{p}} 
-
|\Phi_{\lambda,\tau}^{\mathrm{p}}|^2
)
\langle 
\hat{S}_{\bm{R}'}^z 
\hat{S}_{\bm{R}}^z 
\rangle
-
\Phi_{\lambda,\tau}^{\mathrm{p}\ast} 
\Phi_{\lambda,-\tau}^{\mathrm{p}} 
\langle 
\hat{\bm{S}}_{\bm{R}'}
\cdot
\hat{\bm{S}}_{\bm{R}} 
\rangle
\right]
,
\end{equation}
where $\hat{S}_{\bm{R}}^j=\frac{1}{2}\sum_{\tau,\tau'}\hat{v}_{\bm{R},\tau}^\dagger\sigma_{\tau,\tau'}^j\hat{v}_{\bm{R},\tau'}$ with $j\in\{x,y,z\}$ denotes the VMB spins and $\sigma_{\tau,\tau'}^j$ represents the matrix elements of Pauli operators.
The above expressions connect magnetic correlations to observables in optical experiments, and below we present two ways of extracting $\frac{1}{N}\sum_{\langle\bm{R},\bm{R}'\rangle} \langle \hat{S}_{\bm{R}'}^z \hat{S}_{\bm{R}}^z \rangle$ and $\frac{1}{N}\sum_{\langle\bm{R},\bm{R}'\rangle} \langle \hat{\bm{S}}_{\bm{R}'}\cdot\hat{\bm{S}}_{\bm{R}} \rangle$ without the need of the proportionality factor in Eq.~\eqref{eq:I_lambda_p}.

The first way is to eliminate such proportionality factor by by accessing the paramagnetic state, where both $\langle \hat{S}_{\bm{R}}^z \hat{S}_{\bm{R}'}^z \rangle$ and $\langle \hat{\bm{S}}_{\bm{R}}\cdot\hat{\bm{S}}_{\bm{R}'} \rangle$ are zero.
Denoting the value of $I_\lambda^{\mathrm{p}}$ in such a state as $\mathcal{I}_\lambda^{\mathrm{p}}$, we find:
\begin{equation}
\left(
1
-
\frac{\sum_{\tau} \Phi_{\lambda,\tau}^{{\mathrm{p}}\ast} \Phi_{\lambda,-\tau}^{\mathrm{p}}}{\sum_\tau |\Phi_{\lambda,\tau}^{\mathrm{p}}|^2}
\right)
\sum_{\langle\bm{R},\bm{R}'\rangle} 
\langle 
\hat{S}_{\bm{R}}^z 
\hat{S}_{\bm{R}'}^z 
\rangle
+
\frac{\sum_{\tau} \Phi_{\lambda,\tau}^{{\mathrm{p}}\ast} \Phi_{\lambda,-\tau}^{\mathrm{p}}}{\sum_\tau |\Phi_{\lambda,\tau}^{\mathrm{p}}|^2}
\sum_{\langle\bm{R},\bm{R}'\rangle} 
\langle 
\hat{\bm{S}}_{\bm{R}}
\cdot
\hat{\bm{S}}_{\bm{R}'} 
\rangle
\simeq
\frac{3N}{4}
\delta I_\lambda^{\mathrm{p}}
,\quad
\delta I_\lambda^{\mathrm{p}}
=
1
-
\frac{I_\lambda^{\mathrm{p}}}{\mathcal{I}_\lambda^{\mathrm{p}}}
.
\end{equation}
One can play with the polarizations and extract $\frac{1}{N}\sum_{\langle\bm{R},\bm{R}'\rangle} \langle \hat{S}_{\bm{R}'}^z \hat{S}_{\bm{R}}^z \rangle$ and $\frac{1}{N}\sum_{\langle\bm{R},\bm{R}'\rangle} \langle \hat{\bm{S}}_{\bm{R}'}\cdot\hat{\bm{S}}_{\bm{R}} \rangle$ respectively from the above formula:
\begin{equation}
\label{eq:SS_deltaI}
\sum_{\langle\bm{R},\bm{R}'\rangle} 
\langle \hat{S}_{\bm{R}}^z \hat{S}_{\bm{R}'}^z \rangle
\simeq
\frac{3N}{4}
\delta I_\lambda^{\tau}
,\quad
\sum_{\langle\bm{R},\bm{R}'\rangle} 
\langle \hat{\bm{S}}_{\bm{R}}\cdot\hat{\bm{S}}_{\bm{R}'} \rangle
\simeq
\frac{3N}{4}
\frac{\sum_\tau |\Omega_{\lambda,\tau}|^2}{\sum_{\tau} \Omega_{\lambda,\tau}^\ast \Omega_{\lambda,-\tau}}
\delta I_\lambda^x
+
\left(
1
-
\frac{\sum_\tau |\Omega_{\lambda,\tau}|^2}{\sum_{\tau} \Omega_{\lambda,\tau}^\ast \Omega_{\lambda,-\tau}}
\right)
\sum_{\langle\bm{R},\bm{R}'\rangle} 
\langle \hat{S}_{\bm{R}}^z \hat{S}_{\bm{R}'}^z \rangle
,
\end{equation}
where $\delta I_\lambda^{\tau}$ and $\delta I_\lambda^x$ indicates $\delta I_\lambda^{\mathrm{p}}$ with $\bm{e}_{\mathrm{p}}=\bm{e}_\tau$ and $\bm{e}_{\mathrm{p}}=\bm{e}_x$, respectively.
By setting $\Omega_{1,+} = \Omega_{1,-}$, we arrive at the expressions for $\delta I_1^{\tau}$ and $\delta I_1^x$ in the main text.

Another way is to access three different probe polarizations $\bm{e}_{\mathrm{p}}$.
Plugging the three values of $I_\lambda^{\mathrm{p}}$ into Eq.~\eqref{eq:I_lambda_p} yields three coupled equations with three variables: the proportionality constant, $\frac{1}{N}\sum_{\langle\bm{R},\bm{R}'\rangle} \langle \hat{S}_{\bm{R}'}^z \hat{S}_{\bm{R}}^z \rangle$, and $\frac{1}{N}\sum_{\langle\bm{R},\bm{R}'\rangle} \langle \hat{\bm{S}}_{\bm{R}'}\cdot\hat{\bm{S}}_{\bm{R}} \rangle$.
In principle, the spin correlators can also be obtained by solving these equations.

\subsection{Measurement of the intrinsic magnetic order}
\label{sec:measure_intrinsic_spin_order}

In the previous sections, we discussed how to measure spin correlation \textit{induced} by the drives, which are set off-resonantly ($\omega_1 - \omega_2\neq U$) in order to be compatible with the realization of the $XXZ$ model presented in the main text.
On the other hand, one can instead target at the \textit{intrinsic} magnetic correlation, induced by charge kinetic energy $\hat{T}$ which we neglect throughout the work but (strictly speaking) is present.
For such a purpose, the frequencies of optical sources can be set resonantly, i.e., $\omega_1 - \omega_2 = U$.
More specifically, one can drive one of the two modes and consider absorption of the other (weak) mode.
The absorption rate of such process is from a state in the lower Hubbard band (indicated by the bracket below) to the upper Hubbard band (indicated by the projector $\hat{P}_U$) can therefore be estimated by Fermi-golden rule as $\Gamma_{\mathrm{LU}}\delta(U - \omega_1 + \omega_2)$, where $\Gamma_{\mathrm{LU}}=2\pi\langle\hat{H}_{v}\hat{P}_U\hat{H}_{v}\rangle$.
Note that the terms in the bracket are quartic in VMB electrons, which turns into spin-spin interactions similar to $I_\lambda^{\mathrm{p}}$ discussed in the previous section.
We follow the same strategy by comparing $\Gamma_{\mathrm{LU}}$ with its value in the paramagnetic state, $\Upsilon_{\mathrm{LU}}$ to obtain:
\begin{equation}
\frac{\Gamma_{\mathrm{LU}}}{\Upsilon_{\mathrm{LU}}}
=
1
-
\frac{4}{3N}
\left(
1
-
\frac{
\sum_\tau
\Omega_{2,-\tau}^\ast
\Omega_{1,-\tau}
\Omega_{1,\tau}^\ast
\Omega_{2,\tau}
}{\sum_\tau |\Omega_{1,\tau}\Omega_{2,\tau}|^2}
\right)
\sum_{\langle\bm{R},\bm{R}'\rangle} 
\langle \hat{S}_{\bm{R}}^z \hat{S}_{\bm{R}'}^z \rangle
-
\frac{4}{3N}
\frac{
\sum_\tau
\Omega_{2,-\tau}^\ast
\Omega_{1,-\tau}
\Omega_{1,\tau}^\ast
\Omega_{2,\tau}
}{\sum_\tau |\Omega_{1,\tau}\Omega_{2,\tau}|^2}
\sum_{\langle\bm{R},\bm{R}'\rangle} 
\langle \hat{\bm{S}}_{\bm{R}}\cdot\hat{\bm{S}}_{\bm{R}'} \rangle
.
\end{equation}
Again, by varying the polarization of the drives, one can extract different sectors of spin correlations from the measurement of $\frac{\Gamma_{\mathrm{LU}}}{\Upsilon_{\mathrm{LU}}}$.

\section{High temperature expansion for the Stokes intensities}

In this section, we present the calculation for high temperature expansion on the triangular lattice $XXZ$ model at zero magnetic field:
\begin{equation}
\label{eq:XXZ}
\hat{H}_{J}
=
\hat{H}_{z}
+
\hat{H}_{xy}
,\quad
\hat{H}_{z}
=
\sum_{\langle\bm{R},\bm{R}'\rangle}
J_z
\hat{S}_{\bm{R}}^z
\hat{S}_{\bm{R}'}^z
,\quad
\hat{H}_{xy}
=
J_{xy}
\sum_{\langle\bm{R},\bm{R}'\rangle}
(
\hat{S}_{\bm{R}}^x
\hat{S}_{\bm{R}'}^x
+
\hat{S}_{\bm{R}}^y
\hat{S}_{\bm{R}'}^y
)
,
\end{equation}
with $J_z$ and $J_{xy} = J_z\alpha$ denoting the spin-spin interactions.
We first consider expanding the partition function $Z\equiv\mathrm{Tr}\left[e^{-\beta\hat{H}_{J}}\right]$ in series of inverse temperature $\beta J_z$ up to third order $Z\simeq\sum_{n=0}^3Z_n$.
Each component reads:
\begin{equation}
Z_0 = 2^N
,\quad
Z_1 = 0
,\quad
Z_2 
= 
\frac{\beta^2}{2} 
\mathrm{Tr}\left[
\hat{H}_z^2
+
\hat{H}_{xy}^2
\right]
,\quad
Z_3
=
-
\frac{\beta^3}{6}
\mathrm{Tr}\left[
\hat{H}_z^3
+
3
\hat{H}_z
\hat{H}_{xy}^2
+
\hat{H}_{xy}^3
\right]
.
\end{equation}
To proceed, evaluation of the traces is required.

We start by computing the terms in $Z_2$, which only involves spins in two nearest-neighbor sites, and therefore, produces a $2^{N-2}$ factor for all terms by tracing out other sites.
First, at each site, the spins have to appear as even power as $\mathrm{Tr}[(\hat{\bm{S}}_{\bm{R}})^n] = 0$ with odd $n$.
Together with $\mathrm{Tr}[(\hat{S}_{\bm{R}}^z)^2] = \frac{1}{2}$, we find $\mathrm{Tr}[\hat{H}_z^2] = 2^N\times\frac{3NJ_z^2}{16}$, where we have utilized the fact that triangular lattices have $3N$ bonds in total.
Similarly, with $\mathrm{Tr}[(\hat{S}_{\bm{R}}^x)^2] = \mathrm{Tr}[(\hat{S}_{\bm{R}}^y)^2] = \frac{1}{2}$ and $\mathrm{Tr}[\hat{S}_{\bm{R}}^x\hat{S}_{\bm{R}}^y] = 0$, we obtain $\mathrm{Tr}[\hat{H}_{xy}^2] = 2^N\times\frac{3NJ_{xy}^2}{8}$.
In contrast, terms in $Z_3$ can be classified into two classes.
The ones involving spins at three sites within the same triangular plaquette can be evaluated as  $\mathrm{Tr}[\hat{H}_z^3] = 2^{N}\times\frac{3NJ_z^3}{16}$ and $\mathrm{Tr}[\hat{H}_{xy}^3] = 2^{N}\times\frac{3NJ_{xy}^3}{8}$ following the arguments above.
As for the other terms involving spins in two nearest-neighbor sites, we need $\mathrm{Tr}[\hat{S}_{\bm{R}}^x \hat{S}_{\bm{R}}^y \hat{S}_{\bm{R}}^z] = \frac{i}{4}$, which yields $\mathrm{Tr}[\hat{H}_z\hat{H}_{xy}^2] = -2^{N}\times\frac{3NJ_zJ_{xy}^2}{32}$.
Combining all these ingredients yields: 
\begin{equation}
\frac{Z}{2^N}
=
1
+
N
\left[
\frac{3\beta^2}{32}
(J_z^2+2J_{xy}^2)
-
\frac{\beta^3}{64}
(
2J_z^3
+
4J_{xy}^3
-
3J_zJ_{xy}^2
)
\right]
+
\mathcal{O}(\beta^4J_z^4)
.
\end{equation}
The free energy per site is then:
\begin{equation}
f
=
-\frac{1}{N\beta}
\ln Z
\simeq
-
\frac{\ln2}{\beta}
-
\frac{3\beta}{32}
(J_z^2+2J_{xy}^2)
+
\frac{\beta^2}{64}
(
2J_z^3
+
4J_{xy}^3
-
3J_zJ_{xy}^2
)
+
\mathcal{O}(\beta^3J_z^4)
.
\end{equation}
For the isotropic case where $J_{xy}=J_z=J$, we find $-\beta f \simeq \ln2 + \frac{9(\beta J)^2}{8} + \frac{3 (\beta J)^3}{8}$, consistent with reported results~\cite{oitmaa1996high}.
From the above expression, we can evaluate the following spin correlators at finite temperature:
\begin{equation}
\frac{1}{N}
\sum_{\langle\bm{R},\bm{R}'\rangle} 
\langle 
\hat{S}_{\bm{R}}^z
\hat{S}_{\bm{R}'}^z
\rangle_{\beta}
=
\frac{\partial f}{\partial J_z}
=
-
\frac{3\beta}{16}
J_z
+
\frac{3\beta^2}{64}
(
2J_z^2
-
J_{xy}^2
)
,
\end{equation}
\begin{equation}
\frac{1}{N}
\sum_{\langle\bm{R},\bm{R}'\rangle} 
\left[
\langle 
\hat{\bm{S}}_{\bm{R}}
\cdot
\hat{\bm{S}}_{\bm{R}'} 
\rangle_{\beta}
-
\langle 
\hat{S}_{\bm{R}}^z
\hat{S}_{\bm{R}'}^z
\rangle_{\beta}
\right]
=
\frac{\partial f}{\partial J_{xy}}
=
-
\frac{3\beta}{8}
J_{xy}
+
\frac{3\beta^2}{32}
(
2J_{xy}^2
-
J_zJ_{xy}
)
,
\end{equation}
where $\langle\hat{O}\rangle_{\beta}\equiv\frac{1}{Z} \mathrm{Tr}\left[\hat{O}e^{-\beta\hat{H}_{J}}\right]$ denotes the expectation value of the operator $\hat{O}$ with respect to a thermal state.
The ratio between these correlators can be related to the Stokes intensity via Eq.~\eqref{eq:SS_deltaI}, giving:
\begin{equation}
\frac{\delta I_1^x - \delta I_1^\tau}{\delta I_1^\tau}
=
\alpha
\frac{
2
-
\frac{\beta J_z}{2}
(
2
\alpha
-
1
)
}{
1
-
\frac{\beta J_z}{4}
(
2
-
\alpha^2
)
}
.
\end{equation}
Within the conditions $\Omega_{1,+} = \Omega_{1,-}$ and $\Omega_{2,+} =  e^{i\phi_2} \Omega_{2,-}$, the above formula yields the curves in Fig.~4 in the (End Matter of the) main text.

\bibliography{Biblio}